# Understanding Political Divisiveness using Online Participation data from the 2022 French and Brazilian Presidential Elections

Carlos Navarrete[1], Mariana Macedo[1], Rachael Colley[4], Jingling Zhang[1], Nicole Ferrada[1], Maria Eduarda Mello[6], Rodrigo Lira[7], Carmelo Bastos-Filho[7], Umberto Grandi[4], Jerome Lang[5], César A. Hidalgo[1,2,3,*]

[1] Center for Collective Learning, ANITI, TSE, IAST, IRIT, Université de Toulouse
[2] Alliance Manchester Business School, University of Manchester
[3] Center for Collective Learning, CIAS, Corvinus University
[4] IRIT, Université Toulouse Capitole
[5] PSL, CNRS, Université Paris-Dauphine
[6] Sociology Department, Federal University of Pernambuco, Recife, Pernambuco, Brazil
[7] Computer Engineering Department, University of Pernambuco, Recife, Pernambuco, Brazil
*to whom correspondence should be addressed: cesar.hidalgo@tse-fr.eu & hidalgo.cesar@uni-corvinus.hu

**Abstract**

Digital technologies can augment civic participation by facilitating the expression of detailed political preferences. Yet, digital participation efforts often rely on methods optimized for elections involving a few candidates. Here we present data collected in an online experiment where participants built personalized government programs by combining policies proposed by the candidates of the 2022 French and Brazilian presidential elections. We use this data to explore aggregates complementing those used in social choice theory, finding that a metric of divisiveness, which is uncorrelated with traditional aggregation functions, can identify polarizing proposals. These metrics provide a score for the divisiveness of each proposal that can be estimated in the absence of data on the demographic characteristics of participants and that explains the issues that divide a population. These findings suggest divisiveness metrics can be useful complements to traditional aggregation functions in direct forms of digital participation.

**Introduction**

Digital technologies provide an opportunity to unbundle participation by allowing citizens to express their preferences over fine-grained alternatives. Yet, despite this opportunity, there has been relatively little work exploring the use of digital participation platforms[1–9] to understand citizens' preferences over many alternatives. Here we use the 2022 French and Brazilian presidential elections as an opportunity to help explore this gap by conducting an organic direct democracy experiment. For this experiment, we developed two digital participation platforms (monprogramme2022.org and brazucracia.org) that allowed users to build personalized government programs by combining proposals from the twelve candidates of the 2022 French presidential election and the six candidates of the 2022 Brazilian presidential election. We used this information to explore agreements, using traditional aggregation functions, and disagreements, by constructing a metric of divisiveness. The latter complements the former by distinguishing among similarly ranked proposals. For instance, two proposals ranked 50 and 51 out of 100 alternatives, could rank similarly because participants don't have a strong preference for them, or because some participants strongly support the proposal while others strongly reject it. This difference, which traditional aggregation functions fail to capture, is important to separate tepid proposals, that citizens are relatively indifferent about, from controversial proposals, which are strongly supported or rejected by distinct segments of the population.

In this paper, we show that divisiveness metrics can be constructed as complements for any aggregation function—the aggregates used to identify winners in an election—and provide information about citizens' preferences that is uncorrelated with that provided by its corresponding aggregation function. We use this method to explore the data collected during the 2022 French and Brazilian presidential elections and explore some of its axiomatic properties. To conclude, we use matrix factorization techniques to show that divisiveness is related to higher order eigenvectors of the matrix of pairwise preferences (while aggregation functions are related to the first eigenvectors), suggesting that divisiveness represents a natural extension of aggregation functions for high dimensional data. These findings should be of interest of theorists and practitioners working to design more direct forms of digital civic participation.

**Social Choice and Digital Participation**

Online participation systems have gained prominence in recent years in planned activities, such as participatory budgeting exercises[4,5,10,11], and during spontaneous mobilizations, such as the recent networked social movements[12], in Taiwan[13], Chile[1], and Lebanon[14]. The goal of these digital participation efforts was to provide a more nuanced view of citizen's preferences, and in the case of mobilizations, a means to unlock mobilizations from tactical freeze,[2,15] situations where the social movement is unable to clearly articulate demands.

From an academic perspective, however, digital participation is highly multidisciplinary. On the one hand, its technical design involves challenges in user interface design, privacy, and cybersecurity, as well as mathematical knowledge on social choice theory. On the other hand, the outcomes of digital participation are of the interest of scholars in the social sciences and the humanities, from the political scientists and economists studying the dynamics of attention and polarization, to the scholars exploring the societal impacts of technology. In this paper we contribute to these different streams of literature by using a privacy preserving digital participation platform to explore the creation of aggregates that complement the information provided by traditional aggregation functions and that can extend our understanding of political polarization and attention.

The design of MonProgramme2022.org and Brazucracia.org was inspired by recent crowdsourcing efforts focused on paired comparisons[16–18]. But it deviates from these efforts by using also a fallback voting method[19], where users first agree or disagree with the proposals presented in a list, and then rank-sort the proposals they have accepted (Figure 1 a and b). These designs were chosen to provide a nuanced form of participation, like the one obtained using pairwise comparison platforms, but that could be completed in less time using mobile phones. By having users first approve, and then rank alternatives, we were able to obtain over 1 million preferences from a sample of less than 2,000 participants in France and 150,000 preferences from a sample of nearly 750 participants in Brazil, where we used both approvals and pairwise comparisons.

Participation in MonProgramme2022.org and Brazucracia.org was completely voluntary and organic. Thus, our data does not provide a representative view of the French or Brazilian population, but only of those who participated in it. This means our study is not an attempt to provide a representative view of an electorate, but to explore methods complementing traditional aggregation functions to gain information about the preferences of a given universe of participants. Users in these platforms were allowed to leave at any point or explore the "results" page, where they could find a sorted list of proposals based on their preferences as well as a list of proposals based on the aggregate preferences of all participants (Figure 1 a). After answering about 20 proposals, participants were invited to complete a basic demographic survey (optional), self-reporting information about their gender (M,F,O), location (French department or Brazilian State), level of education, age, political orientation, and the candidate the participant intended to vote for (in Brazil). The protocol for the platform was approved by the Ethics Committee of the Institute of Advances Studies of the Toulouse School of Economics, France (Reference codes #2022-02-001 for France and #2022-07-001) and the University of Pernambuco, Brazil (Reference code CAAE: 61406822.8.0000.5207).

The design of these platforms was privacy preserving, anonymizing personal identifiers (such as IP addresses) using a one-way hash immediately when a user entered the website. To reduce the risk of participants focusing on candidates instead of alternatives, we did not provide information about the candidates associated with each proposal at the time of participation. MonProgramme was publicly released on the morning of March 29 of 2022, two weeks before the first round of the French presidential election and one week after the 12 candidates shared their programs on their official websites or social media. Brazucracia was publicly released on September 28 of 2022, four days before the first round of the Brazilian presidential election. The list of alternatives was curated from the government programs by the research team, which included lawyers, experts in social choice theory, and researchers experienced in the deployment of digital participation efforts.

The mathematical aspects of these online participation platforms connect with the literature on social choice theory. This is an axiomatic theory exploring which voting systems are normatively desirable. While social choice theory can be traced back to the eighteenth century, to the work of the Marquis of Condorcet and Jean-Charles de Borda, a more recent example is Arrow's impossibility theorem[22,23]. Arrow's theorem shows that in elections involving three or more alternatives no ranked electoral system can create a complete and transitive ranking satisfying four key axioms: independence of irrelevant alternatives (IIA) (removing a non-winning candidate does not change the outcome of the election), weak Pareto efficiency (the outcome of the election cannot be improved for every individual), non-dictatorship (no single voter can impose its order of preferences), and unrestricted domain (all possible orderings of preferences are allowed).[24]

At first glance, Arrow's theorem seems like bad news for digital participation efforts. The use of dozens of alternatives implies that it is impossible to run a perfect election satisfying these four axiomatic properties. Yet, the empirical nature of the experiment allows us to ask not if each axiomatic property is perfectly satisfied, but to which extent. Here we follow this goal, since an election that is—for instance—90% robust to the removal of irrelevant alternatives is still desirable over one that is only 50% robust (even though both technically violate the axiomatic property). Moreover, we use this data to explore extensions of traditional aggregation functions capturing information about the divisiveness of each proposal.

By introducing a metric of divisiveness, our results speak also to the literature on political polarization and attention.

During recent years, polarization has become a popular topic of study[25–31]. Yet, despite great interest, a recent survey[26] shows that many studies fail to provide a precise definition of polarization or to distinguish between ideological and affective forms of polarization[25–27]. The former, involves disagreements on ideas and the latter affective responses to opposing political groups, which are often traced to political identities[32,33]. In fact, studies based mostly in the United States, argue that increased political polarization is not a result of changes in ideology, but of the rise of partisan identities within the electorate[25,32,33].

Another important question in political science and media studies involves agenda setting and attention[34–37]. Agenda setting is the ability to direct public attention to a few topics. This research goes back to seminal work showing the ability of the media to focus people's attention on a few topics, even if they do not affect how people think about them[34]. During the last decade, this research was expanded to social media, which provides an opportunity for citizens to set the agenda. In fact, recent research found that topics discussed online by politicians tend to follow those discussed by partisan and politically active citizens, but not general citizens[36], suggesting that politicians tend to respond or follow the agenda of politically active citizens. A related piece of cross-country research also found that polarized discussions tend to increase the perceived relative importance of a topic among citizens.[38] Together, these results suggest that polarized discussion among politically active citizens could have an agenda setting effect.

Our work speaks to both streams of literature.

First, we provide an "atomic" view of ideological forms of polarization by estimating the level of divisiveness of dozens of specific policy proposals (120 in France and 67 in Brazil). But also, we connect our work to political and demographic identities using the self-reported data provided by the participants. In fact, we find divisiveness to be a multidimensional phenomenon with issues related to various self-reported characteristics, from political affiliation (where divisiveness is strong) to self-reported demographic characteristics (age, gender, location). Moreover, by providing a method to estimate divisiveness from participatory behavior, our approach departs from more aggregate methods, such as the use of self-reported left-right Likert scales to study polarization[26]. Finally, our measure of divisiveness is also related to other metrics of polarization used in the economics of conflict literature [39] and in statistical physics[40]. In the case of the former, measures of polarization tend to rely on demographic or economic categories (e.g. income polarization) to estimate population levels of polarization. In our case, we apply these methods using data on self-reported preferences over policy alternatives, allowing us to estimate a level of divisiveness or polarization associated with each alternative. This added level of disaggregation provides a means to understand not only whether a population is polarized, but why.

Second, when it comes to attention, our work shows how we can use a digital participation platform to surface information about the divisiveness of issues that are relevant to a specific group of participants. In a world where politicians follow citizens[36], this could provide a channel to surface information that is relevant to less politically active citizens and goes beyond the issues preferred by more partisan or politically engaged citizens.

In the remainder of the paper, we explore the data collected in these digital participation experiments by first introducing the basic statistics of participation, and then exploring the preference data using traditional aggregation functions and through a metric of divisiveness. Finally, we explore some of the axiomatic properties of social choice theory and matrix factorization techniques, showing that divisiveness represents a mathematically natural extension of traditional aggregation functions.

**Results**

Figures 1 c-p show the number of participants in the platform together with their self-reported demographic information. Participants came mostly from locations with important urban areas (e.g. Paris, Toulouse, Lyon, Bordeaux in France and Sao Paulo and Pernambuco in Brazil), skewed towards the political left, were highly educated (Masters and PhDs), and involved mostly males (72% in France, 61% in Brazil). Since this is clearly not a representative sample of the electorate, we do not present our results as a valid representation of French or Brazilian voters. Instead, we use them to explore the development of methodologies to characterize the divisiveness of the preferences expressed by a specific population of participants.

First, we look at the consistency and transitivity of our sample. Consistency is defined as the fraction of times a participant provided preferences in the same order when having to sort the same pair of proposals together. Transitivity is defined as the fraction of non-cyclical triplets over all observed triplets (see SI for details). Overall, we find the consistency of our sample to be 79.2% (89.2%) and its transitivity to be 82.6% (74.6%) in France (Brazil).

We then study the alignment of the self-reported political orientation of participants and of the candidates from which each proposal originates. Figures 2 a-d show confusion matrices, a standard technique used in computer science to understand classifiers, comparing the self-reported political orientation of participants and the political orientation of the candidates behind each proposal. We classified a proposal as coming from the left or right, if the proposal was present in 50% or more of the candidates labeled as left- or right-wing. This exercise was relatively straight-forward for all candidates except for Emmanuel Macron in France. Thus, in the case of France, we present three matrices, removing Macron from the dataset (Figure 2 a), grouping his proposals with those on the right (Figure 2 b) and on the left (Figure 2 c). In the case of Brazil, we focus on the two main candidates, Lula and Bolsonaro, and consider only proposals that appeared on the government program of only one of these candidates (since some proposals appeared on both programs) (Figure 2d).

The confusion matrices show that left-wing participants tend to agree with proposals from left-wing candidates and right-wing participants tend to agree with proposals from right-wing candidates. In Brazil, this agreement is strong and the off-diagonal elements (e.g. left-wing participants agreeing on right-wing candidate proposals) are fairly symmetric (and rather large). In France however, we observe an important asymmetry, with left-wing participants showing relatively little support for right-wing proposals (<30%) compared to the support for left-wing proposals shown by right-wing participants (>56%). While this result would need to be replicated with a statistically representative sample, it suggests that within our sample of participants in France, right-wing participants were more accepting of left-wing proposals than vice-versa.

Next, we estimate participants' agreement and disagreement with the proposals in the platform. Tables 1 and 2 in the SI rank proposals by their win percentage ($W$): the fraction of times a proposal was selected over other proposals divided by the number of times it was presented to a user (the platforms presented proposals in French and Portuguese, we translated them to English for presentation purposes). This aggregation function is our main measure of agreement and is related to the Borda count (an aggregation function that awards points to a proposal based on the number of other proposals ranked lower than it). The top proposals in France include the "use of 100% renewable energy by 2050," the "increase of personnel in public hospitals," and "increasing the minimum wage." In Brazil, the top ranked proposals were "Valorize the minimum salary to recuperate the purchasing power," "Create a program that expands the guarantee of citizenship for the most vulnerable and brings a universal minimum income," and "Invest on the management of the SUS (the public healthcare system)."

Agreements, however, tell us little about frictions among the population of participants. This motivates us to explore a second form of aggregation focused on disagreements. Our intuition is that divisive proposals, which are strongly supported by one group but opposed by another, can represent opportunities for political trading or bargaining when groups may consider accepting proposals that are low priority for them but important for the opposite group. Divisive proposals can also point to items that need to be deliberated upon or discussed, since they involve issues that the population disagrees on (and hence, may hold different views about). So next, we explore measures of divisiveness, first by leveraging demographic data, and later, by introducing a measure of divisiveness that is agnostic about any self-reported demographic information.

We define the divisiveness $d_i$ of a proposal with respect to a split in the population ($A$ and not $A$ ($\tilde{A}$)) as the difference in the score or ranking $S_i$ that the proposal gets when evaluated in these two sub-populations. Divisiveness is thus defined for any function that maps a set of preferences over alternatives into a numeric value (e.g. Borda, Copeland, etc.). For instance, if a proposal has a win percentage of 60% when evaluated among participants above the median age, and a win percentage of 40% when evaluated among participants below the median age, its divisiveness with respect to median age would be $d(Age)=20\%$. Thus, we define the divisiveness $d$ of proposal $i$ with respect to population $A$ and a score or ranking function $S$ as:

$$d_i(A) = S_i(A) - S_i(\tilde{A}) \qquad (1)$$

Where $S_i(A)$ is the score of proposal $i$ when evaluated in population $A$ and $S_i(\tilde{A})$ is the score of proposal $i$ when evaluated in population not $A$.

Figures 2 e-j show the divisiveness $d$ of proposals in France measured using win percentages when splitting participants according to their self-reported demographic and political orientation data. We explore differences in (e) self-reported political orientation, (f) geography (capital vs regions) and (g) zone (rural vs urban areas), (h) sex, (i) age (top quartile versus bottom quartile), and (j) educational level. When we look at divisiveness by political orientation (e), we find that "Restoration of border control by France leaving the Schengen agreements (ranked 36 overall)", ranks first among participants who identify with the right and center-right ($W = 65.2\%$), but has much less support among participants identifying with the political left ($W = 26.9\%$). This high-level of political divisiveness ($d = 38.3\%$) means the proposal moves 38.3 percentage points when we compare its standing among those who self-identify as members of the political left or right. Similarly, we observe a high level of divisiveness for "Expel foreigners whose behavior is part of radical Islamism and who are registered in the anti-terrorism files" (rank 93, $d = 36.9\%$, $W_{left} = 25.6\%$, $W_{right} = 62.5\%$), and for "Deport foreign offenders at the end of their sentence" (rank 77, $d = 32.8\%$, $W_{left} = 28.4\%$, $W_{right} = 61.2\%$). Conversely, "Create a citizen income", is a proposal that ranks higher among participants who self-identified with the political left, but relatively low among those who self-identify from the right (rank 7, $d = 17.6\%$, $W_{left} = 66.2\%$, $W_{right} = 48.6\%$). Similarly, Figures 2 k-n show the divisiveness $d$ of proposals in Brazil exploring differences in (k) political orientation, (l) geography (here, capital encompasses municipalities of Brasilia, Sao Paulo and Rio de Janeiro), (m) sex, (n) age, and (o) educational level. The proposal to develop "specific health policies aimed at women, LGBTQIA+ population, disabled people and among other vulnerable groups" (rank 31, $d = 37.7\%$, $W_{left} = 58.4\%$, $W_{right} = 20.7\%$) ranks higher among participants who self-identified with the political left, but is less prioritized by participants self-identified from the political right. Conversely, "Expand the privatization of state-owned companies and national infrastructure concessions" (rank 52, $d = 62.9\%$, $W_{left} = 12\%$, $W_{right} = 74.9\%$) is the main priority for participants self-identifying with the political right, but it is ranked low for participants self-identifying with the political left. When we split participants by their self-reported sex, geography (capital versus regions, urban versus rural), age (older and younger people), and education, we observe less divisiveness, but we still identify some divisive proposals that match each dimension. For instance, a proposal to "Reserve social security assistance only for people of French nationality" (rank 40 overall) is more popular among older participants than among younger participants in France; or a proposal for "Equal pay policy between men and women performing the same function" (rank 11 overall) is the second priority for women but the 22$^{nd}$ for men in Brazil. This suggests that, even though political identity might be a leading source of divisiveness, this is nevertheless a multidimensional phenomenon.

At a coarse level, we can measure the overall agreement among two sub-populations by looking at the correlation between the scores obtained by the same proposals $(R^2)^{41}$. Similarly, we can define an aggregate metric of divisiveness $D$ for a population split ($A$ and not $A$) as $(1 - R^2)$. We find in both countries that political orientation provides the largest level of divisiveness ($D_{France} = 1 - R^2 = 69.6\%$, $D_{Brazil} = 1 - R^2 = 100\%$), followed by sex in France ($D_{France} = 1 - R^2 = 27,6\%$) and age in Brazil ($D_{Brazil} = 1 - R^2 = 47.6\%$). Splitting the population by education or location, shows less divisiveness, supporting the literature arguing that polarization is more closely related to political identities, than to demographic identities.[25,27,32,33]

But is there a way to understand divisiveness in absence of self-reported demographic and political orientation data?

**Divisiveness**

Next, we introduce an estimate of divisiveness that is agnostic about any self-reported demographic data. We leverage the fact that each pair of proposals $\{P_i, P_j\}$ divides the population of users into two sub-populations (of not necessarily the same size): those who chose proposal $P_i$ and those who chose proposal $P_j$. We then estimate divisiveness as the average difference in score of a proposal when evaluated among those that selected it, and those that did not, across all pairs of proposals. That is:

$$D_i = \frac{1}{N-1} \sum_{j \neq i} \sqrt{\left(S_i(P_i > P_j) - S_i(P_j > P_i)\right)^2} \qquad (2)$$

Where $(P_i > P_j)$ represent the population of users that selected proposal $P_i$ in the pair $\{P_i, P_j\}$, and $N$ is the total number of proposals. This definition is related to measures of polarization in the literature[39,42–44]. We note that this definition of divisiveness $D$ resembles the definition of standard deviation, and thus, could be interpreted as a second statistical moment for any measure of agreement $S$ (e.g. Copeland, Borda, Elo, AHP, etc.) (in this analogy the score of an aggregation function $S$ represents the first moment or average).

Figure 3 compares the win percentage and divisiveness of proposals, showing that divisiveness is largely uncorrelated with win percentage ($R^2 < 3\%$ *p-value*=0.121 in France; $R^2 \sim 12\%$ *p-value*<0.01 in Brazil). This is in fact a more general property of divisiveness that also holds for other aggregation functions, such as Elo (see SI). The most divisive proposals for participants in France according to these estimate are: "Create a Constituent Assembly to pass to the Sixth Republic" (rank 11 overall, $D = 53.6\%$), "Do not send French soldiers to Ukraine" (rank 89, $D = 53.2\%$), "Engrave in the Constitution the superiority of French law over international law" (rank 38, $D = 51.8\%$), "Restore ENA (the National School of Administration)" (rank 117, $D = 50.8\%$), and "Acquisition of French nationality only by descent or by merit" (rank 27, $D = 50.6\%$). For participants in Brazil, these are "Revocation of the spending ceiling" (rank 41, D=70.7%), "Expand the privatization of state-owned companies and national infrastructure concessions (rank 52, D=67.8%), "Investment in the Armed Forces and promotion of their international participation as in UN-sponsored missions (rank 67,

D=67.3%), "Maintain current labor legislation" (rank 60, D=61%), and "Consolidate and expand land regularization actions, allied to the strengthening of legal institutions that ensure access to firearms" (rank 66, 60.5%). These differences are large (50 percentage points or higher), meaning they can move a proposal from the bottom of the ranking to the top depending on the subset of the population (e.g. from 20% win rate to 70% win rate). Next, we explore the convergence of measures of divisiveness and agreement by estimating the fraction of the sample needed to obtain the ranking obtained with the full sample of France (Figure 3c). Interestingly, we find that divisiveness rankings converge more slowly than rankings of agreements. To achieve the same correlation with the full sample ranking we need only 40,000 preferences for the ranking of agreements (win percentages) but about 300,000 preferences (7.5 times more) for the ranking of divisiveness. This is likely because divisiveness is more of an ensemble property, that unlike agreements, would be non-sensical in a two-candidate election.

Finally, we explore the relationship between the demographic free estimate of divisiveness and divisiveness estimated using self-reported characteristics (figures 3 d-i for France, figures 3 j-n for Brazil). The limited correlations tell us that divisiveness is a multidimensional phenomenon,[45] with each estimate capturing proposals that we can expect to be divisive for each population split. For instance, participants with a master's or PhD education in France are more in support of increasing R&D spending (10) whereas female participants support more strongly the addition of the right to terminate pregnancy in the constitution (27).

**The empirical boundaries of social choice**

To conclude, we explore our data using some of the axiomatic properties of social choice theory. We focus on two key properties: pairwise efficiency and robustness to irrelevant alternatives.

Pairwise efficiency is related to the idea that, in an election, winning alternatives should defeat losing alternatives in a pairwise majority contest. This is inspired by the idea of a Condorcet winner but extended to all pairs of alternatives. We define the pairwise efficiency of an aggregation function as the fraction of times in which a higher-ranked alternative beats a lower-ranked alternative. That is, for a ranking *R*, and a pair of proposals *A* and *B*, with *A* ranked higher than *B*, we count that pair as *pairwise efficient* if *A* defeated *B* in their pairwise contest. The overall pairwise efficiency is defined as the percentage of pairwise efficient pairs.

Figure 4 a illustrates the pairwise efficiency of our data by showing the matrix of all pairwise contests. Rows and columns in this matrix are sorted by win percentage (agreement) and cells represent the head-to-head contest between the proposal in the row and the one in the column. The matrix is characterized by a colored gradient, showing that proposals that are ranked higher tend to beat proposals that are ranked lower. This difference in win percentage increases with differences in ranking (higher ranked proposals tend to beat lower ranked proposals by a larger margin). Aggregating across the whole matrix we find a pairwise efficiency of 81.2% in France and 79.6% in Brazil, meaning that although the ranking is not perfect with respect to pairwise efficiency, it is close.

Next, we study the robustness of the rankings of agreements (win percentage) and disagreements (divisiveness) to the removal of single proposals (independence of irrelevant alternatives or IIA). This is the idea that an election should not change if one of the non-winning alternatives did not run. We focus on two forms of robustness: the top preference (the "election winner") and the robustness of the full ranking (regardless of how high or low a proposal ranked). Figures 4 b-c (France) and k-l (Brazil) present matrices showing the differences in ranking experienced by proposals after removing the proposal represented by the column (sorted by the win percentage). We only count changes in ranking of more than four positions (for other implementations see SI). With this definition we find a robustness of 100% for the top proposal for both agreements and divisiveness. To look at the overall robustness, we count all changes in the ranking, finding a robustness of 87.7% (90.2%) for the ranking of agreements and 33.9% (60.9%) for the ranking of divisiveness in France (Brazil), suggesting that divisiveness may be less robust, but this would require additional research to be established.

Finally, we ask whether divisiveness naturally extends its respective aggregation function by analyzing the matrices of pairwise preferences (Figure 4 a) using singular value decomposition (SVD) (Figures 4 d-f for France, Figures 4 m-o for Brazil). SVD is a fundamental matrix factorization technique that generalizes the concept of eigenvectors and eigenvalues to non-square matrices. SVD can be used to approximate a matrix as a sum of other matrices in a way that it is optimal given a number of factors[46]. Formally, SVD allows us to decompose a matrix $M$ into a sum $M_1 + M_2 + M_3 + \ldots + M_N$, where $M_1$, $M_2$, etc. are outer products of single vectors.

Figures 4 d-f and m-o show the first three factor matrices (e.g. $M_1$, $M_2$, and $M_3$) of the SVD decomposition for France and Brazil, respectively. The first factor is a matrix that is close to 100% pairwise efficient (FR: 95.6%, BR: 97.4%), but not far from the original matrix. This factor accounts for 92.9% of the variation in the original matrix in France (measured as the ratio between the square of its eigenvalue and the sum of the squares of all eigenvalues). This factor is related to the ranking of agreements since it involves a matrix that is virtually free from pairwise violations and comes from an eigenvector that is strongly correlated with the win percentage aggregation function ($R^2$ (France)=97.1%, $R^2$ (Brazil)=98.9%). This means that the information about disagreements must be contained in higher-order factors. In fact, the third eigenvector (accounting for 0.3% of the variation in the matrix) shows a mild but significant correlation with divisiveness in France (Figure 4 i) ($R^2$=19.0%) and the fifth eigenvector correlates with divisiveness in the data from Brazil (Figure 4 r) ($R^2$=13.3%). These findings suggest that divisiveness could be related to higher order factors of pairwise preference matrices and can be thought of as a natural mathematical extension of aggregation functions resulting in scores or rankings (which are related to the first and second eigenvectors).

Finally, to understand whether these properties are peculiar to our sample, or hold more generally, we repeat our analysis using two datasets from preflib.org[47], an academic website specialized in preference data. Figure 5 a-f repeat the analysis for preferences on sushi collected through a survey[48], and judge scores given in Olympic skating free dancing. This show that our two main results hold in very different contexts: people's preferences over sushi and judges' preferences over free skating Olympic dancers. These results are: (i) the lack of correlation between our measures of divisiveness and their corresponding method of aggregation, and (ii) the relationship between the eigenvector decomposition of the pairwise preference matrix and voting rules (connecting the first eigenvector to consensus-based aggregates and high-order eigenvectors to divisiveness). The first result shows that divisiveness metrics bring in information that is new compared to that captured by consensus-based aggregates, and the second one shows that this is related to the spectral decomposition of the matrix of pairwise contests. By showing that these results are valid in data on people's preferences about sushi, judges rating of Olympic athletes, and in the preferences collected through our collaborative government program builder exercise, we demonstrate these two results are unlikely to come from biases in our data, since they hold using complete preference data in very different application domains.

**Discussion**

Here we studied data collected in an online participation platform that allowed participants to build personalized government programs during the 2022 French and Brazilian presidential elections. This sample allowed us to experimentally explore an electoral contest involving 60+ proposals, finding that efforts to understanding preferences in such a context requires complementing traditional measures of consensus with measures of divisiveness.

This is interesting for a variety of reasons.

First, divisiveness provides information that cannot be surfaced through traditional aggregation functions designed to identify the most agreed upon proposals. This is because for every participant that pushes a divisive proposal up in the ranking, there will be other participants pushing it down. The proposals for a new constitution, or for not sending French soldiers to Ukraine, can be clearly identified as the top proposal when we look at the ranking of divisiveness, whereas they would be harder to identify using a ranking of agreements.

There are good reasons to believe that divisive issues are politically relevant. According to some scholars, they provide a better mechanism to signal a person's political position in front of society.[49] For instance, a presidential candidate supporting an increase in the minimum wage—a popular but not very divisive proposal—would fail to signal their political identity at the time of the campaign. Thus, one could argue that during campaign periods, it may be strategic for politicians to focus on divisive issues instead of areas of wide agreement.

Second, divisiveness represents a form of aggregation that has been understudied in social choice theory. This may be because social choice theory generally focuses on elections, whereas divisiveness is not a criterion for election, but more of an input for deliberation. Since our platforms focused on 60+ issues instead of a few candidates, it represented a large enough change of scope and scale from traditional elections to generate the need for a form of aggregation providing complementary information. These higher-order aggregates become more relevant in contests involving dozens or even hundreds of issues, suggesting an avenue of research for social choice theory focused on the axiomatic properties of metrics of divisiveness.

Using online tools for digital participation is a growing phenomenon connected to a rising interest in online civic deliberation[6,9,11,50–55]. Civic participation platforms have been deployed with varying levels of success in Taiwan[2,3,56], Spain[7], Italy[57], and Chile[1]. Yet, there are important risks that need to be considered. Although the platform did not go viral, it suffered cyberattacks (see SI for details). Also, the experience showed that combining approval and ranking provides an efficient option for collecting data (we obtained over 1 million preferences from less than 2,000 participants in France and over 150,000 preferences from less than 750 participants in Brazil). However, the voting system used by this platform can be hard to understand for people used to episodic elections in plurality forms of voting (e.g. first-past-the-post )[58]. The design and development of digital participation tools, therefore, needs to overcome important technical and social challenges before these systems can be more widely adopted. Moreover, the adoption of this platform varied enormously by location. The project in France enjoyed the widest adoption in Haute-Garonne (home of the city of Toulouse, where the platform was created), and less adoption in Île-de-France (Paris). Similarly, the project in Brazil enjoyed the widest adoption in Recife (home of the Brazilian researchers involved in the deploy of Brazucracia). This shows the key role that local networks and traditional media can play to engage users in digital forms of participation.

Finally, there is the question of strategic voting and manipulation[59,60], which may be different for aggregation functions and divisiveness. Traditional aggregation functions can be in fact hard to manipulate[61]. Understanding how to manipulate divisiveness, however, would require further research.

Going forward, we expect to see more work on digital participation efforts and social choice theory, due to the increasing prevalence of decentralized movements and the increased availability of communication and participation technologies. We hope these findings help inspire and inform future research.

**Methods**

All procedures performed in studies involving human participants were in accordance with the ethical standards of the TSE-IAST Review Board for Ethical Standards in Research, under the reference code 2022-02-001 for France and 2022-07-001 for Brazil. In addition, the Brazilian platform was approved by a Brazilian Ethics Review Board: Certificado de Apresentação de Apreciação Ética **CAAE:** 61406822.8.0000.5207. Informed consent was obtained from all individual participants involved in the study.

**Data Availability**

The datasets collected during the current study are deposited in Harvard Dataverse at https://dataverse.harvard.edu/dataset.xhtml?persistentId=doi:10.7910/DVN/8E0EA4. The datasets used for validation of our metric of divisiveness are publicly available on Preflib.org and can be found at https://www.preflib.org/dataset/00014 and https://www.preflib.org/dataset/00006.

**Code Availability**

Data were collected via two digital democracy systems released in Brazil and France preceding their respective 2022 Presidential Elections. The code used to create these platforms is available at: https://github.com/CenterForCollectiveLearning/opencracia. Data analysis was conducted using Python (v.3.10.8), and regression analysis was conducted using R (v.4.2.1). Our pipeline includes the use of Pandas (v.1.5.1), NumPy (v.1.23.4), and SciPy (v.1.9.3). The algorithms implemented divisiveness and aggregation functions are publicly available on Comchoice library at https://github.com/CenterForCollectiveLearning/comchoice/. To further validate our metric of divisiveness, we also collected third-party data from Preflib.org, a comprehensive resource maintained by the Computational Social Choice community.

**Acknowledgements**

This project was supported by: the Artificial and Natural Intelligence Toulouse Institute - Institut 3iA: ANR-19-PI3A-0004, the French National Research Agency (ANR) under grant ANR-17-EURE-0010 (Investissements d'Avenir program), the EUROPEAN RESEARCH EXECUTIVE AGENCY (REA) (https://doi.org/10.3030/101086712, and by the European Lighthouse of AI for Sustainability, HORIZON-CL4-2022-HUMAN-02 project ID:

101120237. The work of UG and RC is supported by ANR JCJC project SCONE (ANR 18-CE23-0009-01). JL work was funded in part by the French government under management of Agence Nationale de la Recherche as part of the "Investissements d'avenir" program, reference ANR-19-P3IA-0001 (PRAIRIE 3IA Institute). The funders had no role in study design, data collection and analysis, decision to publish, or preparation of the manuscript. We acknowledge the graphic design support of Anahide Nahhal for the creation of the MonProgramme platform.

**Author Contributions Statement**

C.N. and C.A.H. contributed to the study conception and design, acquisition of data, data analysis, interpretation of data, and drafting of the manuscript. U.G., J.L., R.C, M.M, N.F., R.L., C.B-F., M.E.M., and J.Z., participated in the creation and diffusion of the platforms and provided comments to improve the manuscript.

**Competing Interests Statement**

The authors declare that they have no known competing financial interests or personal relationships that could have appeared to influence the work reported in this paper.

**Figure Legends/Captions**

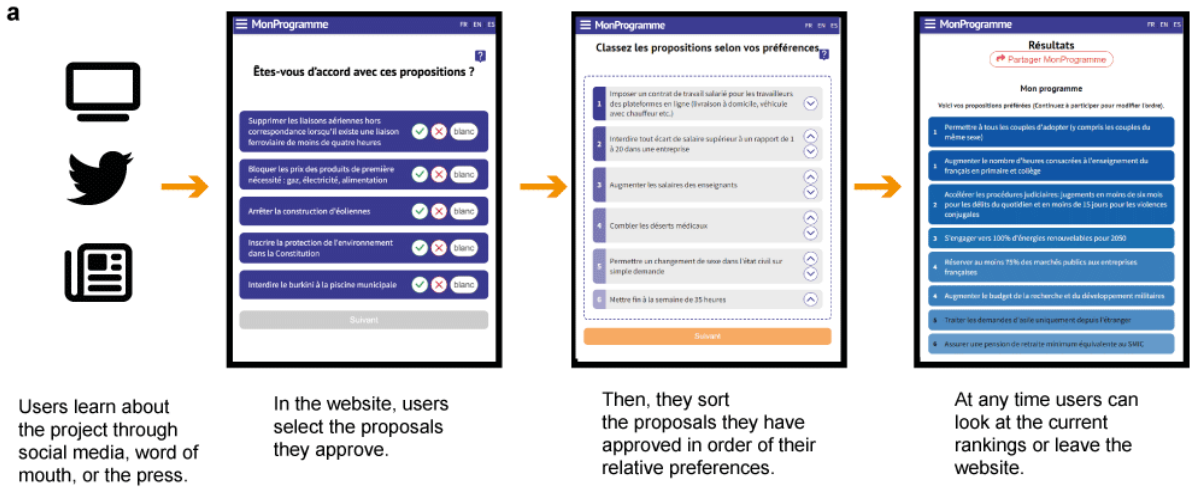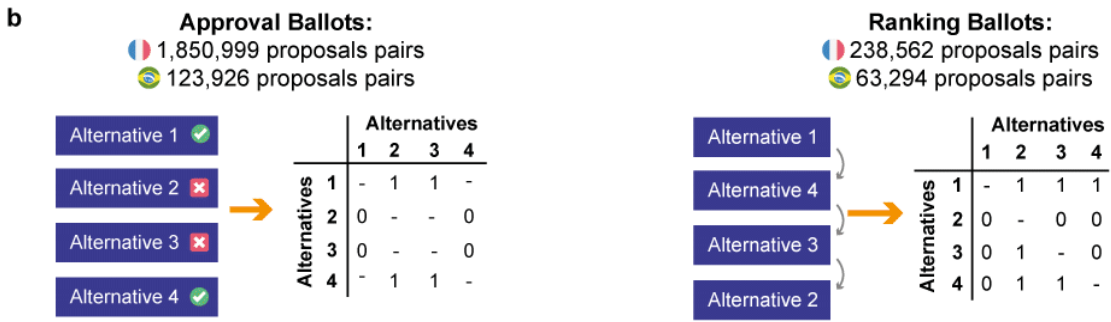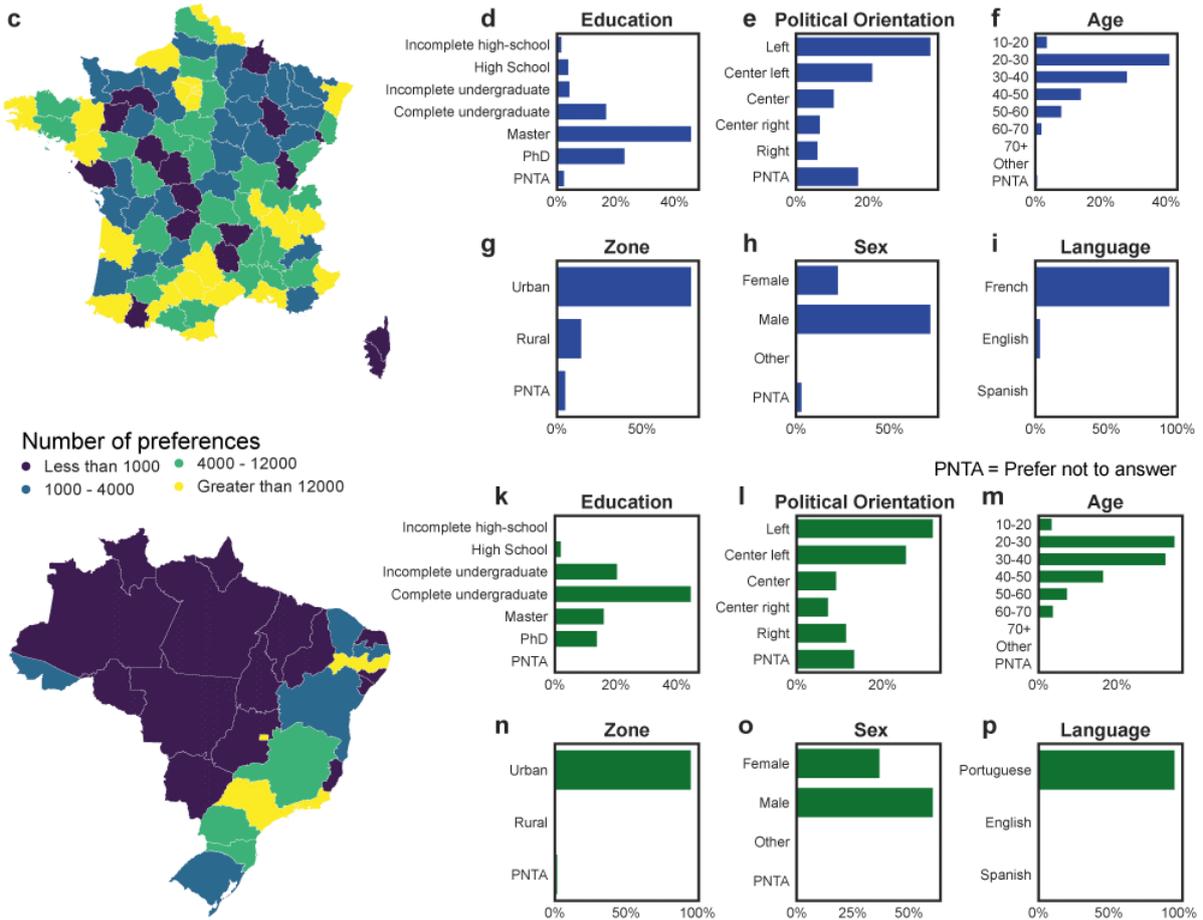

**Figure 1. Participation platform and pairwise comparison data. a.** Basic design of an approval + rank participation platform. Icons were obtained from Font Awesome Free Icons (v.5.0.0) under CC BY 4.0 License. **b.** Procedure to convert approval and rank data into pairwise comparison matrices. Further details of the method in the SI. **c-i** Demographics of digital participation in France ($N_{Participants}$=1175, $N_{Pairs}$=1,705,104) by **c** Geography: France Metropolitan by Departments **d** Education, **e** Political Orientation, **f** Age **g** Zone, **h** Sex, and **i** Language. **k-p** Demographics of digital participation in Brazil ($N_{Participants}$=740, $N_{Pairs}$=157,280) by **k** Geography: Brazilian States **l** Political Orientation, **m** Age, **n** Zone, **o** Sex, and **p** Language. X-axis values represent the percentage of self-reported participants. The number of participants corresponds to the ones that responded to the self-report questionnaire. In case a user has more than one response, we keep the most recent record. Additionally, the number of pairs excludes preference from users targeted as suspicious by our bot detection system. (See SI for details).

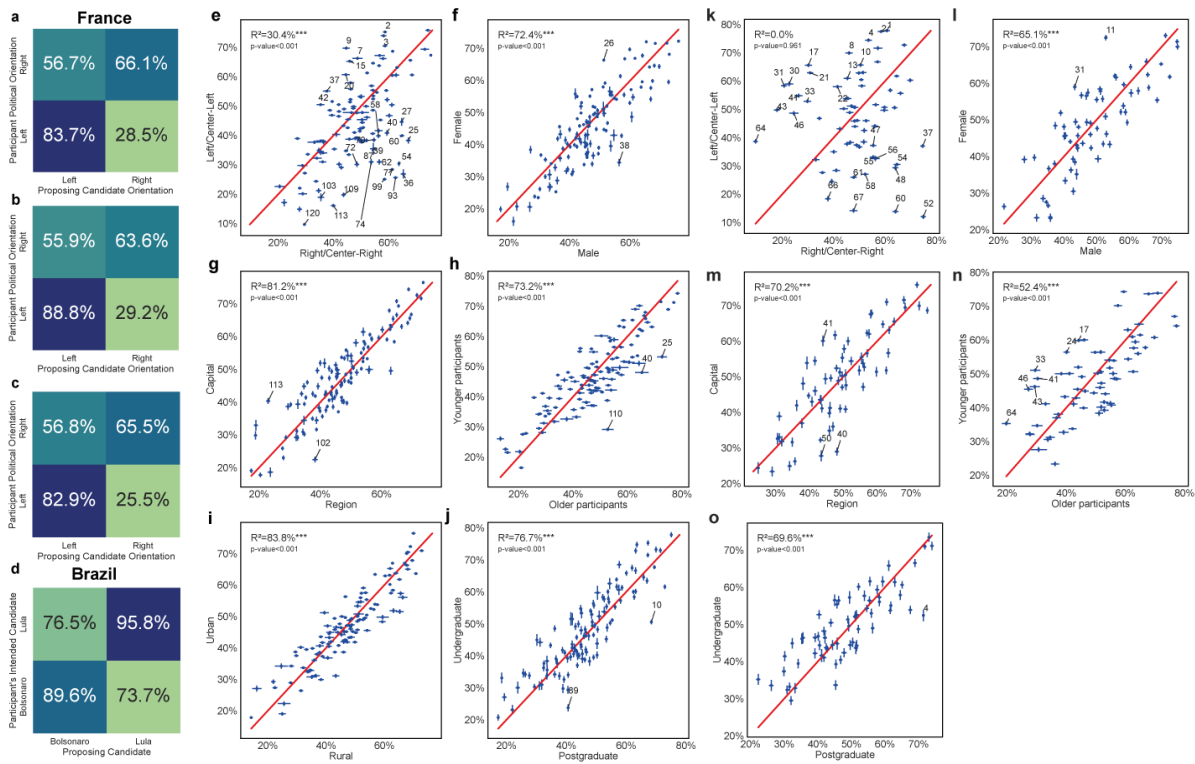

**Figure 2. Understanding participants preferences. a-c** Confusion matrix between participants self-reported preferences and the political orientation of the proposing candidates. ($Participants_{Left}$=690, $Participants_{Right}$=154). Emmanuel Macron was considered as part of the: **a** center (excluded) ($Issues_{Left}$=21, $Issues_{Right}$=9), **b** right ($Issues_{Left}$=19, $Issues_{Right}$=14), and **c** left ($Issues_{Left}$=8, $Issues_{Right}$=11). **d** Confusion matrix between participants intended to vote for Jair Bolsonaro/Lula and the proposals from Bolsonaro and Lula. ($Participants_{Lula}$=346, $Participants_{Bolsonaro}$=113; $Issues_{Lula}$=24, $Issues_{Bolsonaro}$=23). **e-j** comparison of proposal's rankings when splitting the population according to **e** self-reported political orientation ($N_{Left}$=101,190, $N_{Right}$=22,043), **f** location ($N_{Capital}$=35,141, $N_{Region}$=118,067), **g** zone ($N_{Urban}$=131,252, $N_{Rural}$=26,792), **h** sex ($N_{Female}$=43,243, $N_{Male}$=116,697), **i** age ($N_{Younger}$=145,260, $N_{Older}$=20,490), and **j** education level ($N_{Less\ than\ Undergraduate}$ =18,011, $N_{Undergraduate\ or\ more}$ =145,359) in France. **k-n** comparison of proposal's rankings when splitting the population according to **k** self-reported political orientation ($N_{Left}$=24,910, $N_{Right}$=6,470), **l** location ($N_{Capital}$=6,676, $N_{Region}$=32,401), **m** sex ($N_{Female}$=14,437, $N_{Male}$=23,891), **n** age ($N_{Younger}$=33,685, $N_{Older}$=5,243), and **o** education level ($N_{Less\ than\ Undergraduate}$=6859, $N_{Undergraduate\ or\ more}$=31,959) in Brazil. Error bars show 95% CIs computed using 30 bootstrap iterations (score of proposals) of half-size samples and are in some places thinner than the symbols in the figure. We label proposals showing a win percentage difference greater than 15% (IDs correspond to win percentage rank and can be obtained from Supplementary Tables 1

(France) and 2 (Brazil)). For **e-o**, we report the $R^2$ calculated as the square of Pearson's correlation estimated from a two-sided alternative hypothesis, as determined by the SciPy library (v.1.9.3). The number of participants corresponds to the ones that responded to the self-report questionnaire, provided at least one preference (after excluding the "Equal" selection in Pairwise Comparison setup in Brazil), and were not targeted as suspicious by our bot detection system (See SI for details). *Note: \*\*\*p<0.01, \*\*p<0.05, \*p<0.1.*

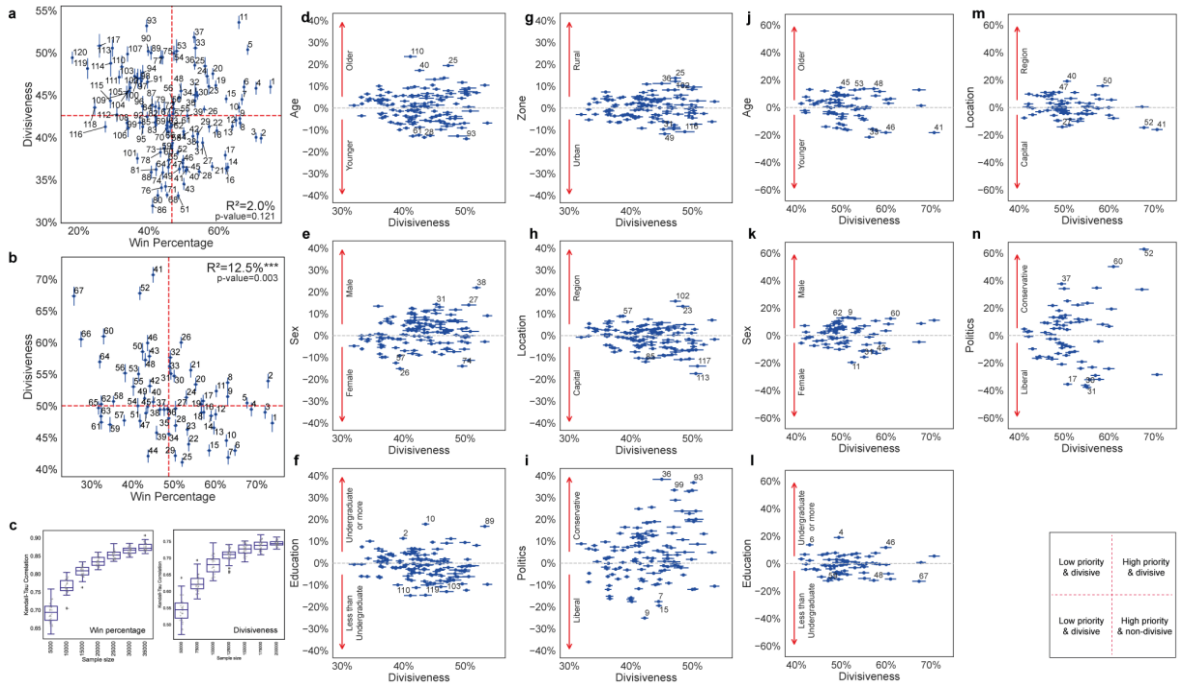

**Figure 3. Divisiveness analysis. Win percentage (x-axis) versus divisiveness (y-axis)** in **a** France (*N*=217,034 pairwise preferences) and **b** Brazil (*N*=49,390 pairwise preferences). For **a-b**, we report the $R^2$ calculated as the square of Pearson's correlation estimated from a two-sided alternative hypothesis, as determined by the SciPy library (v.1.9.3). **c** Convergence in the rankings of agreements and divisiveness estimated using the Kendall-Tau correlation in France. Box plots show the Kendall-Tau correlation between the ranking obtained with the full sample and a random sample of the size indicated in the x-axis. Boxplot figures: center lines show the medians; box limits indicate the 25th and 75th percentiles as determined by the seaborn library (v.0.12.1); whiskers extend 1.5 times the interquartile range from the 25th and 75th percentiles, and circles represent individual data points. **Multidimensional divisiveness** for **d-i** France and **j-n** Brazil. Divisiveness as estimated in equation 2 (x-axis) compared to divisiveness estimated using self-reported (**c (FR), j (BR)**) age, (**e (FR), l (BR)**) education, (**f (FR), m (BR)**) location, (**g (FR), k (BR)**) sex, (**h (FR)** zone, and (**i (FR), n (BR)**) political orientation. Each point in **a-b** and **d-n** represents the mean score of a proposal, and the error bars represent the 95% confidence interval of the proposal's score and are in some places thinner than the symbols in the figure. Both values are calculated by bootstrapping half of the dataset 30 times (See SI for details). *Note: \*\*\*p<0.01, \*\*p<0.05, \*p<0.1.*

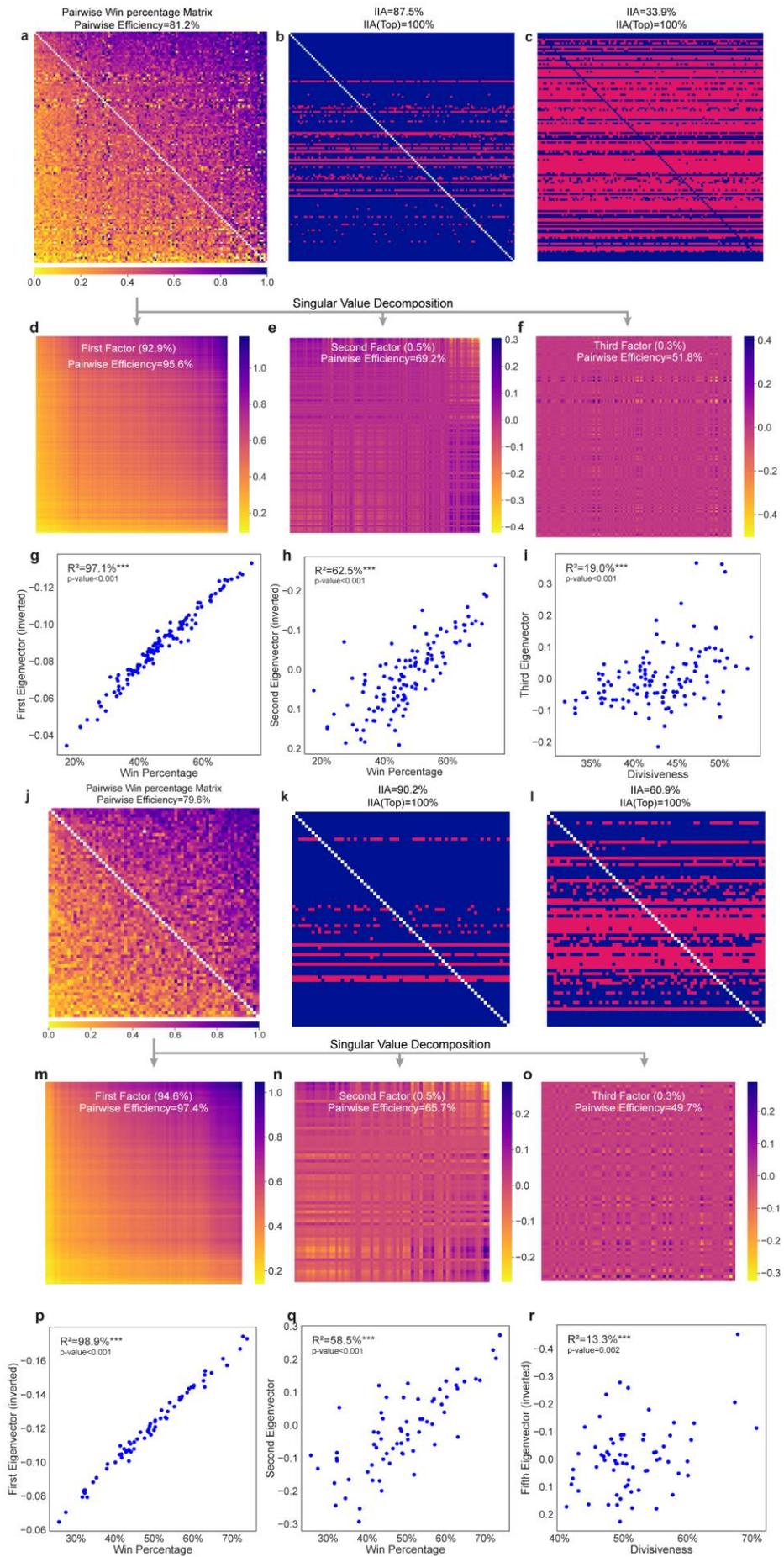

**Figure 4. Axiomatic and matrixial properties of agreement and divisiveness for France (a-i) and Brazil (j-r). (a (FR), j (BR))** Pairwise efficiency of the full matrix of preferences. Rows and columns represent proposals. Values indicate the win percentage of the proposal on the row when competing directly with the proposal on the column. **Independence of Irrelevant Alternatives (IIA)** for **(b (FR), h (BR))** the ranking of agreements (win percentage) and **(c (FR), i (BR))** divisiveness. In (c) and (d) we consider as robust changes involving less than four positions in the ranking (See SI for more details). We computed the ranking of agreements with 30 bootstrap iterations, and no bootstrapping was performed for the ranking of divisiveness. **(FR: d-i, FR: m-o) Singular value decomposition (SVD) of the matrix of pairwise preferences.** Matrices corresponding to the first **(d (FR), m (BR))**, second **(e (FR), n (BR))** and third **(f (FR), o (BR))** factors (eigenvectors). Correlation between first, second and third (fifth BR) eigenvectors (unitary vectors of the Singular Value Decomposition) and **(FR: g-h, BR: p-q)** win percentages and **(i (FR), r (BR))** divisiveness. For **g-i** and **p-r**, we report the $R^2$ calculated as the square of Pearson's correlation estimated from a two-sided alternative hypothesis. *Note: \*\*\*p<0.01, \*\*p<0.05, \*p<0.1.*

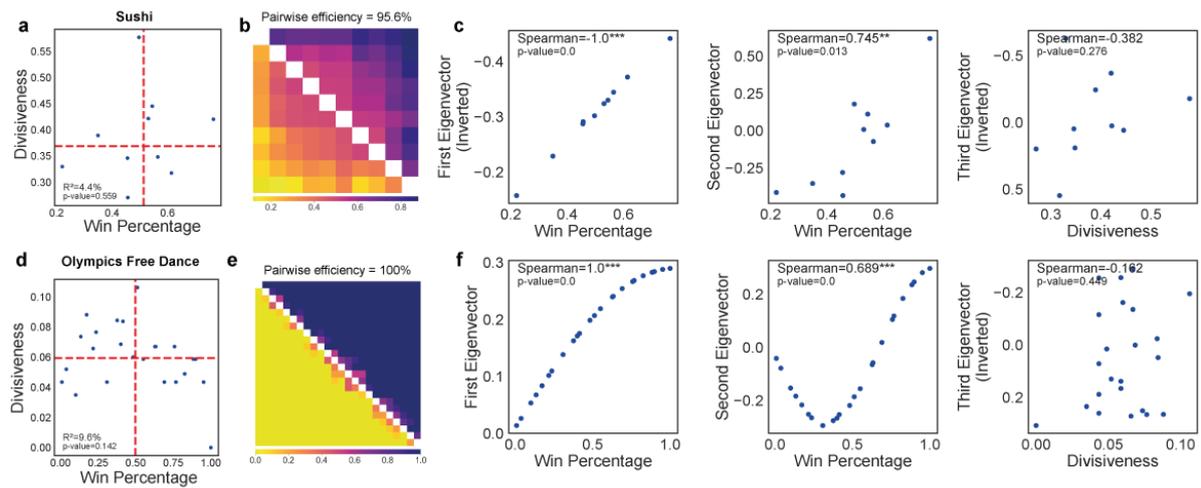

**Figure 5. Reproducing key results in two Strict Order Complete (SOC) datasets stored on Preflib.org**. Win percentage versus divisiveness for **a** Sushi preferences ($Participants_{Sushi}$=5,000, $Proposals_{Sushi}$=10), **d** Judge scoring of free ice-skating dance in the Olympics ($Participants_{Ice\text{-}skating}$=9, $Proposals_{Ice\text{-}skating}$=24), **b-c & e-f** Singular Value Decomposition of these two datasets. For **a, d**, we report the $R^2$ calculated as the square of Pearson's correlation estimated from a two-sided alternative hypothesis, and for **c, f**, we report the Spearman correlation estimated from a two-sided alternative hypothesis. Both values are determined by the SciPy library (v.1.9.3). *Note: \*\*\*p<0.01, \*\*p<0.05, \*p<0.1.*

# Contents

# 1 Supplementary Methods

This section outlines the methods used throughout the manuscript, along with our data processing procedure.

## 1.1 Platform

We collected and curated 120 policy proposals from all candidates in France and 67 from six candidates (Luiz Inácio Lula da Silva, Jair Bolsonaro, Ciro Gomes, Luiz Felipe d'Avila, Soraya Thronicke, Simone Tebet) in Brazil. In each country, this curation work was led by a lawyer. *Mon Programme* was available in French, English, and Spanish; *Brazucracia* was available in Portuguese, English, and Spanish. The proposals' labels included in Tables S1 (France) and S2 (Brazil) represent shorter versions of the ones displayed on the website for clarity.

Our platforms implemented an online version of Fallback Voting (FV) [1]. In the FV method, users first approve or disapprove a subset of proposals and then rank their approved ones. In France, we first show an **approval screen** asking participants to approve, disapprove, or abstain over proposals. Next, we show a **rank screen**, asking them to sort only the ones they previously agreed on. To the best of our knowledge, this is the first time an experimental digital democracy system is based on FV. In Brazil, we slightly modified the experiment by first showing all proposals in repeated **approval screen**, then in a **rank screen**.

This design was chosen to overcome some of the limitations of pairwise comparisons, which in an election of 120 (France) and 67 (Brazil) proposals, involve 7,140 and 2,211 possible pairs to get complete information on relative preferences. Therefore, using FV involves fewer screens than the pairwise comparison, and more data points collected per screen, maximizing our data collection as users could leave the experiment before the end of it.

**Universes**  Both *Mon Programme 2022* and *Brazucracia* asked for individuals' preferences over a large number of proposals. However, asking all 120 proposals in one screen is challenging for users, so we divided the experiment into $k$ proposals per screen, where $k \in \{4, 5, 6\}$. Conversely, we divided the experiment into universes with either 2 or 5 proposals in Brazil. The two-proposals universe was displayed as a classical pairwise comparison screen.

Each user is assigned a random *universe* which is represented by the number of $k$ proposals per screen. Furthermore, they are not made aware of the existence of other universes.

Participation in the platforms was anonymous. We assigned a randomly generated unique UUID per participant, stored as a localStorage() variable in the web browser. The IP address was hashed immediately and automatically using a one-way hash. We also set up Google reCAPTCHA V3 to mitigate the participation of suspicious accounts (we detail more about reCAPTCHA in subsection 1.4).

## 1.2 Datasets

Figure 1 of the manuscript describes the basic design of an approval and rank participation platform. Here, we briefly describe the data collected in each step of the platform and made available for research purposes following the protocol approved by the Ethics Board in France and Brazil.

**Dataset 1: Approval / Disapproval Preferences (Approval Table)** Each row of the approval table represents a participant's approval on a single issue (whether they agrees, disagrees, or does not have a preference). Data collection as of June 6, 2022, of 120,755 rows (France) and as of November 7, 2022 of 25,266 rows (Brazil) stored from the **approval screen** (Table S3).

**Dataset 2: Rank Preferences (Rank Table)** Each row of the rank table represents a participant's ranking of proposals that were displayed to them in a single panel—from the most preferred to the least one. Data collection as of June 6, 2022, of 26,581 rows (France) and as of November 7, 2022 of 20,809 (Brazil) rows from the **rank screen** (Table S4).

**Dataset 3: Pairwise Comparison Preferences** Data collection of pairwise comparison preferences converted from **Dataset 1** and **Dataset 2**, as shown in Figure 1b of the manuscript (Table S5). Each row represents a proposal pair.

First, for each participant in **Dataset 1**, we split the proposals into two groups: the ones she approved and disapproved. Then, we assumed that a participant always prefers a proposal she approved over one she disapproved of, regardless of the screen position. For instance, if a participant agreed with the proposals $\{A, B\}$ and disagreed with proposals $\{C, D\}$, we assumed that she preferred $A$ above $C$, $A$ above $D$, $B$ above $C$, and $B$ above $D$. We omit abstentions from approval screens.

Second, for **Dataset 2**, as each row includes a ranking of proposals, we assume that all preferences in the same panel are transitive. Thus, given $\{A \succ B \succ C \succ D\}$, we obtain $A$ is preferred to $B$, $C$, and $D$, $B$ is preferred to $C$ and $D$, and $C$ is preferred to $D$. It should be noted that we do not assume the transitivity over the participants' preferences when they are given over different panels.

This dataset stores the proposal IDs in the features *option_a* and *option_b*, and the option selected in the feature *selected*. The suspicious participation score, inferred from the reCAPTCHA score assigned to the panel, it is stored on the feature *score*.

We created a unique identifier for the proposal pairs, referred to as *card_id*,[1] by concatenating the *option_a* and *option_b* features, sorted by the one with the lowest ID, to the highest ID. For instance, whether the proposals displayed to a participant are (1, 2) or (2, 1), *card_id* in both cases is 1-2.

---

[1] The number of possible pairwise comparisons, stored in each *card_id* is $\frac{N(N-1)}{2}$, where $N$ corresponds to the number of proposals.

**Dataset 4: Self-Reported Data (Participant Table)** We asked participants to voluntarily self-report six socio-demographic characteristics (sex, age range, political orientation, zone, education, and location) through a popup questionnaire (Table S6).

The popup was displayed automatically after answering a certain number of preferences that varied by each universe (France: 20 for universes 4 and 5; 18 for universe 6—or 5, 4 and 3 panels for universes 4, 5, and 6, respectively; Brazil: 2 panels for universe 2 and 5). Nonetheless, participants could also access the questionnaire manually.

We removed duplicated records associated with the same UUID, keeping the last record (based on **datetime**). Since self-reported data was voluntary, this data set does not include records of all the participants in the platform (Figure S1).

As Figure 2 in the manuscript describes, we could split the participants by their self-reported socio-demographic information. Then, we divide the participants into two groups in each dimension. Table S7 shows the labels used to categorize a participant in one group or another inside each dimension.

**Dataset 5: Set of Proposals** In both campaigns, for each proposal we stored their languages and the candidates who supported them In the French platform we stored proposals in French, English, and Spanish (France) and in the Brazilian platform we stored the proposals in Portuguese, English, and Spanish.

**Dataset 6: Intended Preferences (Intended Table)** In the Brazilian platform, we asked participants to report their intended presidential preference after answering four panels. In total, we collected the intended vote from 550 participants.

### 1.3 Participation

Figure 1a of the manuscript details the participation mechanism of *Mon Programme* and *Brazucracia*. Users must accept the consent form before starting their participation; otherwise, they were redirected to the about page.

After a user completed their participation in the approval screen–that is, after approving or disapproving the set of proposals– the platform then asked them to rank some of their approved proposals. Note that, ordering the same two proposals in different rank screens is an event with a low probability (we address this in detail in Subsection 1.10).

We promoted the platform on social media (e.g., Twitter, Reddit, and Facebook), and a regional newspaper in Toulouse published an article about *MonProgramme*. In Brazil, we advertise the platform using Facebook Ads after the first round. In addition, the universities involved in the project diffused the platform throughout their institutional channels.

## 1.4 Bot Detection System

The interest in distorting public opinion on political issues through bots has been addressed in the literature [2, 3]. This is a common issue for platforms related to measuring citizen preferences [4] or related to candidates during a presidential election.

To mitigate suspicious participation, we implemented reCAPTCHA V3—an API provided by Google, which helps to detect non-human participation by verifying requests that come from a user by returning a score between 0 and 1. Even though reCAPTCHA is not infallible in detecting suspicious actions [5], this score contributes to the purpose of flagging suspicious participation.

We define six criteria (independent of each other) to target participants that exhibited abnormal behaviour regarding the volume or frequency:

1. **Unknown universe**. Participants who voted in a universe not included in the experiment. As we mentioned before, the values accepted for universes are 2, 4, 5, and 6. Nevertheless, we identified users associated to universes not defined in the experiment.

2. **User not registered in the consent form**. We developed front end components to prevent that users participates without accepting the consent form. Nevertheless, we found data points from UUIDs not registered in the consent form table.

3. **Static rank screen**. Participants that did not update proposals in the ranking panel, i.e., change the order of the proposals, in less than 10% of panels. Here we consider users that participated at least three times in ranking panel.

4. **Suspicious accounts detected by Google reCAPTCHA**. Participants with an average reCAPTCHA score of less than 0.7.

5. **Suspicious IP Addresses detected by Google reCAPTCHA**. Participants belonging to IP addresses with an average reCAPTCHA score of less than 0.7.

6. **Over-participation**. Participants that registered preferences in the approval table for more than the maximum number of proposals.

If a participant is part of at least one of these criteria, we consider it a suspicious account. We repeated the procedure for **Dataset 1** and **Dataset 2**. Then, we flagged the suspicious UUIDs detected in both datasets. Following these criteria, we removed 57 suspicious participants in France and 37 in Brazil.

## 1.5 Political Responsiveness in France

Figure 2 a-c of the manuscript shows the percentage of people from the "left" and the "right" of the political spectrum who have preferences towards proposals from candidates with the same or opposite political orientation. To do this, we labeled a proposal as "left" or "right" wing based on the political orientation of the candidates who included it in their government programme. For instance, we say that an issue comes from the left when 50% or more of the candidates labeled as left-wing (Mélenchon, Jadot, Roussel, Hidalgo, Poutou, Arthaud) and less than 50% of the candidates labeled as right-wing (Zemmour, Dupont-Aignan, Pécresse, Le Pen, Lassalle) included it in their government program, and vice versa for the right wing.

In the case of Emmanuel Macron (a well-known "centrist"), we tested three scenarios: (A) excluding Macron from both groups, (B) grouping Macron with right-wing candidates, and (C) grouping Macron with left-wing candidates. Our results were similar across these three specifications.

**Right-wing candidates**  Eric Zemmour, Nicolas Dupont-Aignan, Valérie Pécresse, Jean Lassalle, Marine Le Pen.

**Left-wing candidates**  Anne Hidalgo, Yannick Jadot, Fabien Roussel, Jean-Luc Mélenchon, Phillipe Poutou, Nathalie Arthaud.

## 1.6 Aggregation Functions

We focus our attention on the aggregation functions that use Pairwise Comparison (PC) [6] since we use this data format throughout this manuscript. PC is studied in multiple communities, such as in social choice theory [7], optimization problems, video games, matchmaking, and decision-making process.

Now, we provide a brief summary of four aggregation functions: Win Percentage, Analytic Hierarchy Process, Copeland, and Elo system, along with some applications and their formulas. Consider that we have $m$ issues, and our goal is measuring the score for issue $i$ with a given aggregation method $S$.

### 1.6.1 Win Percentage (W)

The Win Percentage ($W_i$)–or winning percentage of a proposal $i$, is a well-known method to quantify performance in sports competitions that adapted to our purpose, it represents the fraction of times that an issue $i$ was selected. In a pairwise comparison data set, let $x_{ij}$ be the total number of wins of issue $i$ over $j$ in the entire dataset.

The formula to calculate the win percentage for incomplete information is the following [8]:

$$W_i = \frac{\sum_{j=1}^{m} x_{ij}}{\sum_{j=1}^{m} (x_{ij} + x_{ji})} \quad (1)$$

Win percentage values vary between 0 and 1. If a proposal has $W = 0.6$, it indicates that it was selected in 60% of the comparisons.

### 1.6.2 Analytic Hierarchy Process (AHP)

The Analytic Hierarchy Process (AHP) [9, 10] is a decision-making method that compares issues through multiple criteria in order to reach a collective decision. This method is widely used in decision-making processes such as ranking alternatives or quality management [11].

Due to the fact that AHP is rarely studied as a voting rule in social choice, we explain briefly the procedure conducted to calculate the aggregation function using this method [12].

Let $w_{ij}$ the win percentage of issue $i$ over $j$ with the diagonal values $w_{ii} = 1$. Then, we calculate its reciprocal value for the lower triangular matrix as the form $w_{ji} = \frac{1}{w_{ij}}$. The **reciprocal matrix** $A$ as follows:

$$A = \begin{bmatrix} 1 & w_{12} & w_{13} & \cdots \\ \frac{1}{w_{12}} & 1 & w_{23} & \cdots \\ \vdots & \vdots & \ddots & \\ \frac{1}{w_{1m}} & \frac{1}{w_{2m}} & \cdots & 1 \end{bmatrix} \qquad (2)$$

Now, the goal is normalizing $A$ by ensuring that the summation of each column should be 1. For this, we calculate the total of each column from $A$ as follows:

$$X_j = \sum_{i=1}^{m} A_{ij} \qquad (3)$$

Now, the matrix $A'$ in Equation 4 represents a column-normalized version of $A$, where each element is divided by the summation of its column ($X_j$).

$$A' = \begin{bmatrix} 1 & \frac{x_{12}}{X_2} & \frac{x_{13}}{X_3} & \cdots \\ \frac{1}{x_{12}X_1} & 1 & \frac{x_{23}}{X_3} & \cdots \\ \vdots & \vdots & \ddots & \\ \frac{1}{x_{1m}X_1} & \frac{1}{x_{2m}X_2} & \cdots & 1 \end{bmatrix} \qquad (4)$$

We now obtain the principal eigenvector from $A'$. In this literature, it is also called **priority vector**. This vector shows the relative importance of each issue. The summation of the priority vector is 1. For instance, if the priority vector for 3 issues is {0.7, 0.2, 0.1}, we say that the the importance of first issue is 70%.

### 1.6.3 Copeland

The Copeland method [13] ranks issues based on the number of pairwise majority contests an issue wins, loses, and ties with respect to the other issues. Let $x_{ij}$ the number of total wins of issue $i$ over $j$, $m$ the number of issues, and $y_{ij}$ represents whether $i$ beats $j$.

$$y_{ij} = \begin{cases} 1 & \text{if } x_{ij} > x_{ji} \\ 1/2 & \text{if } x_{ij} = x_{ji} \\ 0 & \text{if } x_{ji} > x_{ij} \end{cases} \qquad (5)$$

Next, the Copeland score of an issue $i$ is given by:

$$\text{Copeland}_i = \frac{\sum\limits_{j=1, i \neq j}^{m} y_{ij}}{m - 1} \qquad (6)$$

Copeland ranges between 0 and 1. Consider a dataset without ties, if a proposal has $Copeland = 0.6$, it indicates that it beats 60% of other individual issues. If there is a Condorcet winner (a given issue who beats each other in individual comparisons), the Copeland score for that issue will be 1.

### 1.6.4 Elo system

To measure players' performance and to ensure fair matches, the Elo rating system aims to quantify the relative skill of a player. Originally proposed to quantify the performance of chess players, Elo has inspired research on aggregation functions using this technique [14]. Similarly to AHP, the Elo system is rarely studied as a voting rule.

To compute Elo, first, we set up an equal initial score ($S_0$) for every issue. Then, let $A$ and $B$ be two issues. Consider that issue $A$ has a score $R_A$ and issue $B$ has a score $R_B$, then the expected score for $A$ ($E_A$) and $B$ ($E_B$) are:

$$E_A = \frac{1}{1 + 10^{(R_B - R_A)/S_0}} \qquad (7)$$

$$E_B = \frac{1}{1 + 10^{(R_A - R_B)/S_0}} \qquad (8)$$

$E_A$ and $E_B$ values vary between 0 and 1, and $E_A + E_B = 1$. When $E_A > E_B$, we say that the probability of choosing $A$ over $B$ is greater than the opposite. Then, given that a participant compared the issues $A$ and $B$, the new score for $A$ ($R'_A$) is given by:

$$R'_A = R_A + K(S_A - E_A) \qquad (9)$$

Where $K$ represents a freedom degree in the equation and $S_A$ the score obtained from the comparison (0 in case that $B > A$ and 1 whether $A > B$).

When using the Elo ranking, we set $K = 10$ and $S_0 = 400$. As in this system the order of the rows plays a role in the final score of a proposal, we add another step that shuffles the dataset similar to our bootstrapping procedure (more details in Subsection 1.8).

### 1.6.5  TrueSkill

TrueSkill [14] is a matching algorithm–inspired by Elo rating–developed by the Microsoft team for creating teams to play against one another during video games on Xbox live (with teams of the form 1v1 or NvN). We use the function *trueskill*, implemented using its Python library [15], as an aggregation function that returns a rank based on scores it would use in the matching. In this regard, we input a dataset with pairwise comparisons and the selected options, and the output is the score of proposals. This study uses the Python library trueskill (v0.4.5).

## 1.7  Data Curation

The findings presented in the manuscript were obtained using **Dataset 2**. That is, we use data that come from the **rank screen** to calculate the different ranking of agreements.

This decision is motivated by several reasons. First, **Dataset 2** implicitly contains information about approvals since we only asked to rank issues approved before by participants. Second, given that our method for measuring divisiveness splits the participants into subpopulations, we cannot compute this measure just using pairs from **Dataset 1**. This is due to highly approved or rejected proposals being unable to split the population into groups in contrast to **Dataset 2** in which we collect relative preferences. Third, using a mix of both datasets would bias the result and interpretation of divisiveness in favor of the approval data since **Dataset 1** represents around 90% of the proposal pairs in France. We addressed this point for ranking agreements in Subsection 2.1.3.

The data curation procedure consisted in removing data points using two criteria to reduce possible distortions in the results:

- Participants labeled as suspicious by the Bot Detection System introduced in subsection 1.4.

- Duplicated preferences of a participant on the same pair of proposals. We only kept the last data point based on the **datetime**.

This curated dataset is the one used to perform the analysis presented in the manuscript. Here we will refer to the ranking of agreements (win percentage) presented in the manuscript as **R1**.

## 1.8  Bootstrapping

Given an aggregation function $S$ that is based on a scoring function (Win Percentage, Divisiveness, Elo, AHP), we bootstrapped the data set 30 times to estimate the confidence interval of scores. In each iteration, we randomly sampled half of the dataset. Finally, we calculated the rating of each proposal – as the average score obtained by $S$ in each iteration.

## 1.9 Proposal IDs

After collecting all of the preferences over the proposals, we then relabelled their IDs in the manuscript and SM such that they corresponded to their position in the agreement ranking, calculated by the win percentage.

## 1.10 Consistency and Transitivity

Our platforms allowed users to compare the same two proposals in repeated rank screens. For validity purposes, we tested two measures, namely consistency and transitivity on **Dataset 2**. Both analyses presented in the manuscript are focused on non-suspicious participants.

**Consistency**  We measure the consistency as the fraction of times a user provided the same order for two proposals in repeated rank screens. From a rank of the form $A > B > C$, we assume that $A > B$, $B > C$, and $A > C$. This index is calculated uniquely for proposal pairs compared more than once by the same user. As a side note, we exclude ties reported in universe 2 of pairwise comparisons in the Brazilian data.

**Transitivity**  The transitivity of the elicited preferences is measured as the percentage of three proposals which are not cyclic. That is, for the triplet $ABC$, the possible pairs are $AB$, $AC$, and $BC$ and we want to ensure that the preferences returned by a single user do not lead to a situation where $A > B > C > A$. The procedure we use to quantify the transitivity is as follows:

- We compute all the possible triplets that could have occurred in **Dataset 2**. In the case of our French platform, the number of triplets for 120 proposals is $280,840$ (i.e., $\binom{120}{3}$). In the Brazilian platform, the number of triplets for 67 proposals is $47,905$.

- For each participant, we were interested in the transitive (contiguous) triples answered by them in the same panel from **Dataset 2**. For example, if the participant ranked $A > B > C > D$, the triplets removed were $A - B - C$, $A - B - D$, $B - C - D$.

- In **Dataset 3**, for each contiguous triplet returned by a user, we calculate the percentage of times it is cyclic. For example, consider a participant that has selected $A > B$ 5 times, $B > C$ 5 times, $A > C$ once and $C > A$ once. It is clear that this has been made cyclic with the pairwise comparison over $A$ and $C$. We can find 50 triplets for this participant ($A > B > C$ 25 times and $B > C > A$ 15 times). Therefore, the transitivity of this participant on this observed triplet is 50%.

- We then take the average over all of these percentages calculated from the contiguous triples of each user.

## 1.11 PrefLib Preferences

We extend the analysis presented in the manuscript beyond the data collected by the platform. Thus, we downloaded data from [PrefLib.org](PrefLib.org) [16], an open source repository for researchers interested in testing on preferences, such as computational social choice, and recommender systems.

We downloaded data labeled as *Strict Order Complete (SOC)*, which represents the type of preferences in which voters provided a complete and strict ordering over the alternatives (thus, no ties are allowed between alternatives). This category provides complete data about voters over a set of alternatives, unlike our data collected, which encompasses incomplete preferences.

Now, we describe the PrefLib data sets analyzed and included as part of Figure 5 in the manuscript. Each data set is associated to a unique ID which can be used to find the data set in their repository.

### 1.11.1 Sushi Rank (00014)

This data set was generated from a questionnaire survey of which sushi the users preferred via a ranking method. It includes the preferences from 1000 participants over 10 kinds of sushi (File ID: **00014-00000001**) [17].

### 1.11.2 Skate Data (00006)

This data set contains figure skating rankings from various competitions during the 1998 season including the World Juniors, World Championships, and the Olympics [18]. These data sets generally have 10-25 candidates (skaters) and 8-10 judges (voters).

Here, we analyzed the 1998 Nagaro Olympics Dance Free Dance (File ID: **00006-00000018**), that includes preferences of 24 alternatives and 9 voters.

## 2 Supplementary Results

In this section, we present additional results to the ones presented in the manuscript.

### 2.1 Robustness

In this subsection, we perform a set of robustness analyses for ranking of agreements and divisiveness. First, we compare the ranking of agreements by using different aggregation functions, then by splitting the data set by universe size, data source, and finally, we analyze the robustness of divisiveness.

#### 2.1.1 Robustness by Aggregation Function

We compare the ranking of agreements (Win Percentage) presented in the manuscript (**R1**) with four aggregation functions: Elo (**R2**), AHP (**R3**), Copeland (**R4**), and TrueSkill (**R5**) (Figures S2 and S3). We observed a strong positive Kendall-Tau (KT) correlation between the ranking from these aggregation functions and **R1** (France: Copeland=0.882, Elo=0.927, AHP=0.83, TrueSkill=0.929, Brazil: Copeland=0.94, Elo=0.965, AHP=0.956, TrueSkill=0.969). Our analysis suggests that the ranking of agreements does not vary significantly by comparing with other aggregation functions.

Then, we analyze whether participants labeled as suspicious (probable bots) could distort the results. For this, we compare **R1** with a ranking of agreements (**R6**) that include both suspicious and non-suspicious participation.

We find a strong positive Kendall-Tau correlation (France: $KT = 0.952$, Brazil: $KT = 0.937$) between **R1** and **R6**, suggesting that suspicious activity (probable bots) did not alter the results (Figure S4).

#### 2.1.2 Robustness by Universe

We split **Dataset 2** by universe to compare their aggregated rankings of agreement using Win Percentage. As shown in Figures S5 and S6, overall, we observe in all the subsets a strong positive KT correlation of their rankings (France: $KT > 0.7$, p-value $< 0.01$, Brazil: $KT = 0.619$, p-value $< 0.01$). We point out that most variations are concentrated in the middle of the ranking.

#### 2.1.3 Robustness by Data Source

Our pairwise comparison dataset is generated by the data from Approval Voting (AV) and from the Ranking Voting (RV). Now, we calculate the ranking of agreements using different subsets: Approval, Approval & Rank (the combination of both data sets), and Rank (the ranking data introduced in the manuscript). We observe in Figures S7 and S8 that the ranking of agreements using approval or rank data changes significantly (France: $KT =$

0.3, p-value < 0.01, Brazil: $KT = 0.379$, p-value < 0.01). This can be explained since the approval data collect absolute judgments over the proposals, and the rank data contain relative preferences over a set of issues. For instance, the proposal "Plan to use 100% of renewable energy" in France is ranked 1st in approval but $12^{\text{th}}$ in rank. This could be interpreted as the participants highly approving the proposal, but they did not rank it highly in their ranking.

### 2.1.4 Robustness Divisiveness

To complement the robustness results in the manuscript, we want to show robustness with another aggregation function. Thus, we focus on Elo, showing similar results to Win Percentage. We focus on Elo as in preliminary research, we noticed that Copeland requires more proposal pairs to converge into the expected pattern.

We do not observe a Kendall-Tau correlation between the ranking of divisiveness (introduced in the manuscript) and the one calculated with Elo (France: $KT = -0.086$, Brazil: $KT = -0.012$) (Figure S9). Also, we visualize divisiveness (calculated with Elo) vs. Elo rating, finding that divisiveness (Elo) is slightly correlated to its agreement measure (Figures S10 and S11).

## 2.2 Multidimensional Divisiveness

Figure 3d-i (France) and 3j-n (Brazil) of the manuscript visualizes the divisiveness of each proposal per dimension. Here, in order to enrich that analysis, we included Figures S12, S13, S14, S15, and S16 that visualizes the divisiveness of each proposal by socio-demographic characteristic in France. Each axis represents a dimension (e.g., sex, location, education). Similarly, S17, S18, and S19 present the same analysis for Brazil.

Here we set up a regression model in order to analyze the factors that explain divisiveness. Let $D_i$ be divisiveness (as defined in Equation 2 of the manuscript), $W_i$ be win percentage of a proposal $i$, and $|d_i(X)|$ the absolute value of disagreement of a proposal $i$ with respect to dimension $X$ (Political Orientation (Politics), Sex, Location (Capital vs Region), Age, Education and Zone (Urban vs Rural)) as control variables. We regress divisiveness ($D_i$) as a multidimensional approach of a proposal using the following extended model presented in Equation 10:

$$D_i = \beta_0 |W_i| + \beta_1 |d_i(\text{Politics})| + \beta_2 |d_i(\text{Politics})| + \\ \beta_3 |d_i(\text{Sex})| + \beta_4 |d_i(\text{Location})| + \beta_5 |d_i(\text{Age})| + \\ \beta_6 |d_i(\text{Zone})| + \beta_7 |d_i(\text{Education})| + \epsilon_i \qquad (10)$$

All the model variables were standardized with a standard deviation equal to 1.

Tables S9 (France) and S10 (Brazil) summarizes four regression analyses, one for each socio-demographic factor separately and one controlling by all the factors in France and Brazil. We observe that Location and Politics are positively associated to divisiveness in all the models for France; whereas Politics, Sex and Age are positively associated to divisiveness in Brazil. This finding suggests that divisiveness is a multidimensional component and cannot be explained by just one dimension. For instance, the questionnaire popup did not include a question for self-reported religious beliefs, voting intention, or confidence in institutions.

## 2.3 SVD Decomposition and Divisiveness

We explore whether the eigenvectors from the singular value decomposition (SVD) could help to explain the measures of agreements and divisiveness introduced in the manuscript. We set up linear regressions of the form:

$$y_i = \beta_0 + \beta_1^T \text{Eig.1}_i + \beta_2^T \text{Eig.2}_i + \beta_3^T \text{Eig.3}_i + \epsilon_i \qquad (11)$$

where $y_i$ is the dependent variable for the proposal $i$ (win percentage, divisiveness), $Eig.X$ is the $X$-th eigenvector depicted from the pairwise matrix, $\beta_0$ is the intercept, and $\epsilon_i$ is the error term.

Tables S11 (Dependent variable: Win Percentage, dataset: France), S12 (Dependent variable: divisiveness, dataset: France), S13 (Dependent variable: Win Percentage, dataset: Brazil), and S14 (Dependent variable: divisiveness, dataset: Brazil) summarize the findings of a set of ordinary least squares (OLS) regression analyses to study the relationship between the eigenvectors from the regression and the scores of agreements or divisiveness of the proposals. Overall, we observe that the first and second eigenvectors positively correlate with the measure of agreement. Indeed, those two factors are significant in all the models. Conversely, the third eigenvector seems to have a strong association with divisiveness in France. In Brazil, we again observe that the first and second eigenvector are associated with agreements, but we do not observe associations for the third eigenvector with divisiveness.

## 2.4 Convergence of Agreements and Divisiveness

We present in Figure 3c of the manuscript the convergence for ranking of agreements and ranking of divisiveness in France. Here, we estimate the convergence for Win Percentage and divisiveness scores using the same dataset. To compute the convergence of agreements, we compared the Win Percentage scores from the full data and the Win Percentage on the sampled data. We sampled the data 30 times to calculate confidence intervals on different sizes. Next, we calculate the Kendall-Tau (KT) correlation of each sample. Our goal is determining the minimum number of preferences necessary to reach a threshold of 0.75 KT with between the scores of the partial and full dataset.

We focus on the French data, since the pairwise comparison dataset (**Dataset 3**) is 4.34 times larger than the Brazilian data and the analysis of convergence could be biased due to the sample size. The number of proposals pairs in the Brazilian data is around 50,000, this much lower than the threshold identified from the French data for the convergence of divisiveness.

We observe similar results in for the convergence of scores in Figure S20 to those presented in Figure 3c of the manuscript. We obtained a Kendall-Tau correlation higher than 0.75 with at least 10,000 proposal pairs for agreements and at least 200,000 proposal pairs for divisiveness (around 20 times more than the data necessary to get convergence in the ranking of proposals).

These results suggest that a ranking of agreements using a pairwise comparison paradigm converges with a relatively small set of preferences ($\sim 4 \times 10^5$) with a stricter convergence criterion, and it is more computationally complex to measure divisiveness than agreements with the method proposed in the manuscript.

## 2.5 Independence of Irrelevant Alternatives (IIA)

Let $R$ be a ranking of proposals (either by agreement or divisiveness). Consider that we remove a proposal $i$ from the data set. Then, we calculate a new ranking of proposals $R'$. For each different proposal $j$, we compute the distance $s$ between the ranking of $j$ in the two rankings $R$ and $R'$. To properly compare the rankings, we remove $i$ from $R$ and re-calculate the position of all the proposals in the ranking.

Next, we calculate the percentage of violations by moving a threshold of acceptable distance in the ranking. For instance, suppose we have $R = \{a \succ b \succ c \succ d\}$, and we want to compare with $R' = \{a \succ d \succ c\}$ that is obtained by removing $b$ from data. In this case, we compare it to $\{a \succ c \succ d\}$, a reduction of $R$. The distance in this example is $a = 0$, $c = 1$, $d = 1$.

Figures S21 and S22 present the results when allowing for a distance of zero to seven, both when the ranking the proposals in terms of agreements and divisiveness in France (similarly, Figures S23 and S24 for Brazil). By being less strict in the allowed distance, we increase the percentage of satisfaction.

For instance, by allowing a maximum distance of 5 in the French data, 98.6% of the time, the data satisfies IIA. Similarly, by allowing a maximum distance of 5 in the Brazilian data, the data satisfies IIA 93.4% percent of the time. Furthermore, by looking at the previously mentioned figures, we observe that most of the violations are located in the middle of the aggregated rankings for both divisiveness and agreement.

We will openly share the data collected and curated in this manuscript in CSV format. The data set will also have a version formatted for Preflib [16].

# 3 Supplementary Tables

| Id | Name | Win percentage |
|---|---|---|
| 1 | Plan to use 100% renewable energies by 2050 | 75.1% |
| 2 | Increase personnel in public hospitals | 72.2% |
| 3 | Increase the minimum wage | 71.5% |
| 4 | Further develop the French nuclear park | 70.8% |
| 5 | Include ecology in the Constitution | 68.9% |
| 6 | Reduce working hours to 32 hours per week | 67.6% |
| 7 | Create a citizen income | 66.8% |
| 8 | "Cap prices of essential products: gas, electricity, food" | 66.5% |
| 9 | Prohibit any salary difference of more than 1 to 20 in a company | 66.2% |
| 10 | Devote 3% of GDP to research and development | 65.4% |
| 11 | Create a Constituent Assembly to pass to the Sixth Republic | 65.4% |
| 12 | Lower retirement age to 60 | 65.0% |
| 13 | Increase the industrialization of the country | 63.9% |
| 14 | Ensure a minimum pension is equivalent to the minimum wage | 62.7% |
| 15 | Restore the solidarity tax on wealth (ISF) | 62.5% |
| 16 | Increase number of doctors in rural underserved areas | 62.3% |
| 17 | Increase teacher salaries | 62.2% |
| 18 | Ban dangerous pesticides (eg neonicotinoides) | 59.6% |
| 19 | Nationalize or renationalize some large companies (EG Telecom / Orange) | 59.1% |
| 20 | Increase social housing | 58.5% |
| 21 | Establish a Citizens' Initiative Referendum | 58.3% |
| 22 | End the 35-hour working week | 57.7% |
| 23 | Increase the retirement age | 57.6% |
| 24 | Abolish fee-for-service pricing in hospitals | 56.6% |
| 25 | Abolition of the law of the soil | 56.4% |
| 26 | Add to the Constitution the right to voluntary termination of pregnancy | 55.5% |
| 27 | Acquisition of French nationality only by descent or by merit | 55.4% |
| 28 | Make those convicted of corruption unable to run for office | 54.9% |
| 29 | Index pensions to inflation | 54.8% |
| 30 | Develop a taxation to discourage programmed obsolescence | 54.4% |
| 31 | Elect the National Assembly by proportional voting | 53.7% |
| 32 | Prohibit single-use plastics | 53.7% |
| 33 | Lower VAT on fuels | 53.5% |
| 34 | Oppose intensive breeding and slaughter | 53.3% |
| 35 | Recognition of the blank vote | 53.2% |
| 36 | Restoration of border control by France leaving the Schengen agreements | 53.1% |
| 37 | Guarantee the rights of asylum for refugees | 52.9% |
| 38 | Engrave in the Constitution the superiority of French law over international law | 52.5% |
| 39 | Only process asylum applications from abroad | 51.9% |
| 40 | Reserve social security assistance only for people of French nationality | 50.8% |
| 41 | Invest to maintain and rebuild local railway lines | 50.3% |
| 42 | Increase the amount of APL (housing assistance) | 50.1% |
| 43 | Increase paternity leave to match maternity leave | 49.9% |
| 44 | "Introduce a universal rental guarantee to facilitate access to housing, financed by landlords and the State" | 49.8% |
| 45 | Create police units and magistrates dedicated to violence against women | 49.7% |
| 46 | Reduce legal proceedings time for everyday offenses and domestic violence | 49.6% |
| 47 | Relocate the production of medicines to France | 49.6% |
| 48 | Eliminate air routes when there is a rail connections taking less than 4 hours | 49.6% |

| 49 | Continue the pension reform of the outgoing government | 49.5% |
| --- | --- | --- |
| 50 | Create spaces in EHPAD (accommodation institutions for dependent senior) and recruit staff | 48.8% |
| 51 | Ban facial recognition in public spaces | 48.6% |
| 52 | Introduce a tax on airline tickets when there is a comparable rail route | 48.2% |
| 53 | Impose employment contracts for the workers of online platforms | 48.0% |
| 54 | Introduce immigration quotas by profession and country | 47.2% |
| 55 | Lift patents for Covid-19 vaccines | 46.9% |
| 56 | "Allow all couples to adopt (married or not, including same-sex couples)" | 46.6% |
| 57 | Reimburse medically assisted procreation for all women | 46.5% |
| 58 | Require activities that promote employability to obtain the RSA (income support) | 46.4% |
| 59 | Regulate undocumented migrants | 46.3% |
| 60 | At least 75% of public contracts to be reserved for French companies | 46.3% |
| 61 | Nationalization of the highways | 46.0% |
| 62 | Suspend social benefits of parents of juvenile offenders | 45.9% |
| 63 | Make taxes individual by removing marital quotient | 45.9% |
| 64 | Increase the defense budget by at least 2% of GDP | 45.9% |
| 65 | Eliminate university admission requirements | 45.8% |
| 66 | Increase the military's research and development budget | 45.5% |
| 67 | Establish a mandatory national civic service | 45.5% |
| 68 | Pay adult disability benefits independenly of their spouse's revenue | 45.2% |
| 69 | Increase availability of space at day cares | 45.0% |
| 70 | Create a large number of police and militarized police positions | 44.6% |
| 71 | Leave NATO's Defence Planning Committee | 44.5% |
| 72 | Disenfranchise those who attack those responsible for public authority | 44.5% |
| 73 | Increase number of hours spent in French language education | 44.4% |
| 74 | Limit family reunification | 44.1% |
| 75 | Decriminalization of assisted death | 43.6% |
| 76 | Teach a second language from primary school | 43.5% |
| 77 | Deport foreign offenders at the end of their sentence | 43.3% |
| 78 | Redirect a portion of a city's public funds to support rural areas | 43.2% |
| 79 | Propose a simpler and more advantageous combination of employment and retirement | 42.9% |
| 80 | Departmental elected officials will be the same as regional elected officials | 42.9% |
| 81 | Pay a significant child allowance from the first child | 42.7% |
| 82 | Generalize the teaching of computer code and digital uses from the 5th | 42.7% |
| 83 | Submit foreign investments to the approval of a High Council for Economic and Digital Sovereignty | 42.3% |
| 84 | Pay family allowance independently of family resources | 42.2% |
| 85 | Lower charges for the self-employed | 42.0% |
| 86 | "Limit inheritance tax up to €150,000 per child" | 41.6% |
| 87 | Total corporate tax exemption for entrepreneurs under 30 years old | 41.2% |
| 88 | Refrain from any military intervention without the mandate of the United Nations | 40.3% |
| 89 | Do not send French soldiers to Ukraine | 39.8% |
| 90 | Create a sex-offender registry | 39.7% |
| 91 | Send weapons to Ukraine | 39.7% |
| 92 | Offer student loans | 39.7% |
| 93 | Expel foreigners whose behavior is part of radical Islamism and registered in the anti-terrorism files | 39.1% |
| 94 | Reintroduce a seven-year presidential term | 38.6% |
| 95 | Allow universities to have complete autonomy | 38.1% |
| 96 | Raise the salaries of gendarmes and police officers | 37.9% |
| 97 | Give 16 year olds the right to vote | 37.4% |
| 98 | Restore the life sentences | 36.8% |
| 99 | Close any mosque where sermons are hostile to France and the values of the Republic | 36.6% |
| 100 | Allow non-Europeans to vote in local elections | 36.5% |
| 101 | Lower the age of criminal responsibility from 18 to 16 years old | 36.4% |
| 102 | Oppose Ukraine's entry into NATO | 35.0% |
| 103 | Privatize the audiovisual public service | 34.2% |
| 104 | Replace Parcoursup with a more transparent system | 33.7% |
| 105 | Stop the construction of wind turbines | 33.7% |
| 106 | Reimburse the transition of trans people | 33.6% |
| 107 | Allow a change of sex in the civil status by a simple request | 33.5% |
| 108 | Legalise cannabis | 33.2% |
| 109 | Use facial recognition at the entrance to public transport to better identify wanted people | 33.1% |
| 110 | Reduce the tax on real estate wealth by exempting it from 50% of the main residence | 32.4% |
| 111 | Forbid hunting | 31.1% |
| 112 | Add 2 more hours of sport per week in school | 29.8% |
| 113 | Prohibit the wearing of the veil by school chaperones | 29.7% |
| 114 | Remove of the TV licence fee | 28.1% |
| 115 | Defend regional languages and cultures | 27.6% |
| 116 | Guarantee the free movement of cars in cities | 27.1% |
| 117 | Restore ENA (the National School of Administration) | 23.9% |
| 118 | Establish full autonomy for Corsica | 21.8% |
| 119 | Protect hunting and fishing rights | 21.7% |
| 120 | Prohibit the burkini at municipal swimming pools | 17.5% |

Table S1: List of proposals included in the French platform, ranked by win percentage.

| Id | Name | Win percentage |
|---|---|---|
| 1 | Valorize the minimum salary to recuperate the purchasing power | 73.7% |
| 2 | Create a program that expands the guarantee of citizenship for the most vulnerable and brings a universal minimum income | 72.7% |
| 3 | Invest on the management of the SUS | 72.0% |
| 4 | Strengthening career plans and valuing teachers | 68.6% |
| 5 | Tax reform with change in burden reducing taxation on consumption and increasing income progressively so that the richest pay more | 67.6% |
| 6 | Actions aimed at training and qualification of teachers | 64.7% |
| 7 | Commitment to the goals stipulated by the National Education Plan | 63.0% |
| 8 | Invest in the national system to promote technological development through funds and public agencies such as CNPq and CAPES | 62.9% |
| 9 | No income tax for workers making up to 5 minimum wages. | 62.9% |
| 10 | Strengthening the national vaccination program | 62.6% |
| 11 | Equal pay policy between men and women performing the same function | 60.2% |
| 12 | Invest in specific programs and actions aimed at the educational recovery of those affected by the pandemic | 60.0% |
| 13 | Strengthen the popular pharmacy program | 59.6% |
| 14 | Strengthen the energy supply with the expansion of clean and renewable sources | 58.9% |
| 15 | Actions that guarantee internet access in public schools | 58.4% |
| 16 | Increase transparency through compliance with the Access to Information Law | 57.2% |

| 17 | Continue the policy of social and racial quotas for admission to higher education | 57.0% |
| 18 | Investing in vocational education in line with labor market expectations | 56.6% |
| 19 | Write a new labour legislation to include modern labour regulations and social protection | 56.5% |
| 20 | Propose rules for the transparency of final beneficiaries of public resources | 55.2% |
| 21 | "Strengthen democratic, secular and inclusive education with specific policies for people with disabilities, the LGBTQIA+ population and among other vulnerable groups" | 54.0% |
| 22 | Increase the participation of women in politics and public management | 53.5% |
| 23 | Reduce agricultural production costs and marketing price | 53.2% |
| 24 | Recover lands deteriorated by predatory activities and reforestation of devastated areas | 52.9% |
| 25 | Encouraging sustainable agricultural practices | 51.9% |
| 26 | The complete opening of banking and fiscal secrecy of first and second level positions in the Executive Power. | 51.7% |
| 27 | Curb drug mining and money laundering in the Amazon by increasing the number of ecological bases | 50.3% |
| 28 | Policies and actions for debt renegotiation of households and companies | 50.3% |
| 29 | Encourage research related to medicines | 50.2% |
| 30 | Preservation of culture and demarcation of indigenous and quilombolas lands | 50.0% |
| 31 | "Specific health policies aimed at women, LGBTQIA+ population, disabled people and among other vulnerable groups" | 49.2% |
| 32 | Maintain the value of 600 reais for Auxílio Brasil | 49.0% |
| 33 | Meet the carbon gas reduction targets assumed by the country at the 2015 Paris Conference | 48.9% |
| 34 | Improve and reduce the prices of transport services through the structuring of concessions and public-private partnerships | 48.7% |
| 35 | Structuring the medical career in the SUS with mechanisms of attraction and recognition | 48.6% |
| 36 | Regulation and protection of workers' labor rights by application | 48.4% |
| 37 | Encouraging entrepreneurship through credit facilitation and debureaucratization | 47.6% |
| 38 | Health facilities with good performance should monitor and assist those with lower performance | 46.5% |
| 39 | Creation of policies to provide hybrid work and home office for women with children | 45.7% |
| 40 | Actions to curb tax evasion | 44.9% |
| 41 | Revocation of the spending ceiling | 44.9% |
| 42 | Improve public job positions and salary plans with incentives related to goals | 44.1% |
| 43 | Creation of policies that guarantee the inclusion and permanence of the LGBTQIA+ population in the labor market | 44.0% |
| 44 | Encouraging female entrepreneurship through the facilitation of credit and microcredit | 43.6% |
| 45 | "Combining face-to-face teaching with distance learning in basic education, analyzing regional peculiarities" | 43.5% |
| 46 | Policy for valuing state-owned companies and those against privatization | 43.5% |
| 47 | Implement a federal Public Integrity strategy | 43.1% |

| 48 | "Establish the basic foundations of the subjects, removing ideological connotations and with a view to parents as the main actors in children's education" | 43.0% |
| 49 | Actions to combat illegal mining | 42.8% |
| 50 | New fuel pricing policy | 42.3% |
| 51 | Actions to encourage the creative economy | 41.7% |
| 52 | Expand the privatization of state-owned companies and national infrastructure concessions | 41.7% |
| 53 | Reinforce the consolidation of the national cancer care support program | 41.3% |
| 54 | Expand the articulation between the public and private health sectors | 41.2% |
| 55 | Fostering agro-industry and national production of inputs | 40.1% |
| 56 | Country's formal adherence to the OECD Council's Public Integrity Recommendation | 38.1% |
| 57 | "Offer Green Bonds to finance investments considered sustainable in the areas of transport, energy and between others" | 37.9% |
| 58 | "Expand, redesign, and improve the qualification programs of the police" | 35.3% |
| 59 | Culture-focused policies through articulation with private sector institutions and companies and civil society organizations | 34.4% |
| 60 | Maintain current labor legislation | 32.9% |
| 61 | Implement national guidelines for the promotion and defense of police human rights | 32.3% |
| 62 | Increase national production of fertilizers | 32.3% |
| 63 | Continue programs related to encouraging physical activity for primary care | 32.2% |
| 64 | Create a new National LGBTI+ Public Policy Committee | 31.9% |
| 65 | Encouraging mining activity within a logic of environmental protection | 31.6% |
| 66 | "Consolidate and expand land regularization actions, allied to the strengthening of legal institutions that ensure access to firearms" | 27.4% |
| 67 | Investment in the Armed Forces and promotion of their international participation as in UN-sponsored missions | 25.7% |

Table S2: List of proposals included in the Brazilian platform, ranked by win percentage.

| Column | Description | Type |
| --- | --- | --- |
| id | Auto increment ID | integer |
| user_id | Unique Participant ID | string |
| proposal_id | Proposal ID | integer |
| agree | User's preference about a proposal. Values are approved (1), disapproved (-1) or absent (0) | integer |
| multichoice | Additional information about a proposal approved | string |
| universe | Universe size. Values are 4, 5, 6 | integer |
| score | Google reCAPTCHA V3 score | float |
| created_at | Datetime when data was stored | datetime |
| locale | Language platform | string |

Table S3: Features in Dataset 1.

| Column | Description | Type |
| --- | --- | --- |
| id | Auto increment ID | integer |
| user_id | Unique Participant ID | string |
| rank | Sorted ranking of alternatives displayed in a panel | string |
| updated | Whether a user updated the randomly generated ranking displayed | boolean |
| universe | Universe size. Values are 4, 5, 6 | integer |
| score | Google reCAPTCHA V3 score | float |
| locale | Platform language | string |
| created_at | Datetime when data was stored | datetime |

Table S4: Features in Dataset 2.

| Column | Description | Type |
| --- | --- | --- |
| id | Auto increment ID | integer |
| user_id | Unique participant ID | string |
| option_a | Proposal ID displayed in the first part of the screen. (Left in desktop and top in mobile) | integer |
| option_b | Proposal ID displayed in the second part of the screen. (Right in desktop and bottom in mobile) | integer |
| option_a_sorted | Proposal ID of lowest value between option_a and option_b | integer |
| option_b_sorted | Proposal ID of highest value between option_a and option_b | integer |
| card_id | Unique identifier of a proposal pair | integer |
| selected | Proposal ID selected. If the value is equal to 0 means that the participant selected "Don't have preference" option | integer |
| created_at | Datetime registered | datetime |
| score | Score assigned to the participant by Google ReCaptcha V3. Values are $[0, 1[$, and values close to 0 indicates that there is more probability of an anomalous account | float |
| universe | Universe Size. Values are 4, 5, 6 | integer |
| source | Source of data. Values are "agree" and "rank" | string |

Table S5: Features in the pairwise comparison data set of preferences.

| Column | Description | Type |
| --- | --- | --- |
| id | Auto increment ID | integer |
| user_id | Unique participant ID | string |
| politica | Political Orientation | integer |
| location | Department ID | integer |
| age | Age Range | integer |
| sex | Sex ID | integer |
| education | Education ID | integer |
| universe | Universe Size. Values are 4, 5, 6. | integer |
| created_at | Datetime when data was stored | datetime |
| score | Google reCAPTCHA V3 score | float |
| locale | Platform language. Values are "fr", "en", and "es" | string |

Table S6: Features in Dataset 4. Labels description included in data shared.

| Dimension | Type | Labels |
|---|---|---|
| Political Orientation | Conservative | 4, 5 |
| | Liberal | 1, 2 |
| | Excluded | 5 |
| Location | Region | 1, 2, 3, 4, 5, 6, 7, 8, 9, 10, 11, 12, 13, 14, 16, 17, 18, 21, 22, 24, 25, 26, 27, 28, 29, 30, 31, 32, 33, 34, 35, 37, 38, 39, 41, 42, 44, 45, 46, 47, 49, 50, 51, 53, 54, 55, 56, 57, 58, 59, 60, 62, 63, 64, 65, 66, 67, 68, 69, 70, 71, 73, 74, 76, 79, 80, 81, 82, 83, 84, 85, 86, 87, 88, 89, 90, 972, 973 |
| | Capital | 75, 77, 78, 91, 92, 93, 94, 95 |
| | Excluded | 998, 999 |
| Sex | Female | 1 |
| | Male | 2 |
| | Excluded | 98, 99 |
| Age | Young participants | 1, 2, 3, 4 |
| | Old participants | 5, 6, 7 |
| | Excluded | 99, 98 |
| Education | Less than Undergraduate | 1, 2, 3 |
| | Undergraduate or More | 4, 5, 6, 7 |
| | Excluded | 99 |
| Zone | Urban | 1 |
| | Rural | 2 |
| | Excluded | 99 |

Table S7: Labels and categories used in France for each socio-demographic dimension.

| Dimension | Type | Labels |
|---|---|---|
| Political Orientation | Conservative | 4, 5 |
| | Liberal | 1, 2 |
| | Excluded | 5 |
| Location | Region | All other labels |
| | Capital | 3516 (Rio de Janeiro), 2699 (Sao Paulo), 2392 (Brasilia) |
| | Excluded | 998, 999 |
| Sex | Female | 1 |
| | Male | 2 |
| | Excluded | 98, 99 |
| Age | Young participants | 1, 2, 3, 4 |
| | Old participants | 5, 6, 7 |
| | Excluded | 99, 98 |
| Education | Less than Undergraduate | 1, 2, 3 |
| | Undergraduate or More | 4, 5, 6, 7 |
| | Excluded | 99 |
| Zone | Urban | 1 |
| | Rural | 2 |
| | Excluded | 99 |

Table S8: Labels and categories used in Brazil for each socio-demographic dimension.

|  | Dependent variable: | | | |
|---|---|---|---|---|
|  | Divisiveness | | | |
|  | (1) | (2) | (3) | (4) |
| Win Percentage |  |  |  | −0.087 |
|  |  |  |  | p = 0.315 |
| Age |  |  |  | 0.111 |
|  |  |  |  | p = 0.228 |
| Education |  |  |  | 0.132 |
|  |  |  |  | p = 0.134 |
| Location | 0.270*** |  | 0.216** | 0.195** |
|  | p = 0.003 |  | p = 0.016 | p = 0.031 |
| Politics |  | 0.302*** | 0.256*** | 0.165* |
|  |  | p = 0.001 | p = 0.005 | p = 0.073 |
| Sex |  |  |  | 0.106 |
|  |  |  |  | p = 0.219 |
| Zone |  |  |  | 0.144 |
|  |  |  |  | p = 0.107 |
| Constant | 0.000 | 0.000 | 0.000 | 0.000 |
|  | p = 1.000 | p = 1.000 | p = 1.000 | p = 1.000 |
| Observations | 120 | 120 | 120 | 120 |
| $R^2$ | 0.073 | 0.091 | 0.136 | 0.217 |
| Adjusted $R^2$ | 0.065 | 0.083 | 0.121 | 0.168 |
| Residual Std. Error | 0.967 (df = 118) | 0.957 (df = 118) | 0.938 (df = 117) | 0.912 (df = 112) |
| F Statistic | 9.301*** (df = 1; 118) | 11.810*** (df = 1; 118) | 9.176*** (df = 2; 117) | 4.436*** (df = 7; 112) |
| Note: | | | | *p<0.1; **p<0.05; ***p<0.01 |

Table S9: OLS summary of Divisiveness decomposed by dimension in France.

|  | Dependent variable: | | | |
|---|---|---|---|---|
|  | Divisiveness | | | |
|  | (1) | (2) | (3) | (4) |
| Win Percentage |  |  |  | −0.199** |
|  |  |  |  | p = 0.039 |
| Age |  |  |  | 0.254*** |
|  |  |  |  | p = 0.007 |
| Education |  |  |  | 0.048 |
|  |  |  |  | p = 0.602 |
| Location | 0.146 |  | 0.155 | 0.134 |
|  | p = 0.238 |  | p = 0.114 | p = 0.145 |
| Politics |  | 0.616*** | 0.618*** | 0.478*** |
|  |  | p = 0.00000 | p = 0.00000 | p = 0.00001 |
| Sex |  |  |  | 0.202** |
|  |  |  |  | p = 0.041 |
| Zone |  |  |  | 0.037 |
|  |  |  |  | p = 0.697 |
| Constant | 0.000 | 0.000 | 0.000 | 0.000 |
|  | p = 1.000 | p = 1.000 | p = 1.000 | p = 1.000 |
| Observations | 67 | 67 | 67 | 67 |
| $R^2$ | 0.021 | 0.379 | 0.403 | 0.540 |
| Adjusted $R^2$ | 0.006 | 0.370 | 0.385 | 0.485 |
| Residual Std. Error | 0.997 (df = 65) | 0.794 (df = 65) | 0.784 (df = 64) | 0.718 (df = 59) |
| F Statistic | 1.419 (df = 1; 65) | 39.712*** (df = 1; 65) | 21.628*** (df = 2; 64) | 9.878*** (df = 7; 59) |

Note: *p<0.1; **p<0.05; ***p<0.01

**Table S10:** OLS summary of Divisiveness decomposed by dimension in Brazil. *Note: *p<0.1; **p<0.05; ***p<0.01.*

|  | Dependent variable: | | | | | |
|---|---|---|---|---|---|---|
|  | Win Percentage | | | | | |
|  | (1) | (2) | (3) | (4) | (5) | (6) |
| First Eigenvector | −0.989*** | | −0.984*** | | −0.989*** | −0.984*** |
|  | p = 0.000 | | p = 0.000 | | p = 0.000 | p = 0.000 |
| Second Eigenvector | | −0.693*** | −0.007 | | | −0.007 |
|  | | p = 0.000 | p = 0.803 | | | p = 0.804 |
| Third Eigenvector | | | | −0.252** | 0.001 | −0.001 |
|  | | | | p = 0.043 | p = 0.974 | p = 0.972 |
| Constant | −0.000 | −0.000 | −0.000 | −0.000 | −0.000 | −0.000 |
|  | p = 1.000 | p = 1.000 | p = 1.000 | p = 1.000 | p = 1.000 | p = 1.000 |
| Observations | 65 | 65 | 65 | 65 | 65 | 65 |
| $R^2$ | 0.977 | 0.480 | 0.977 | 0.064 | 0.977 | 0.977 |
| Adjusted $R^2$ | 0.977 | 0.472 | 0.977 | 0.049 | 0.977 | 0.976 |
| Residual Std. Error | 0.152 (df = 63) | 0.727 (df = 63) | 0.153 (df = 62) | 0.975 (df = 63) | 0.153 (df = 62) | 0.154 (df = 61) |
| F Statistic | 2,722.041*** (df = 1; 63) | 58.231*** (df = 1; 63) | 1,340.812*** (df = 2; 62) | 4.274** (df = 1; 63) | 1,339.441*** (df = 2; 62) | 879.476*** (df = 3; 61) |

| Note: | *p<0.1; **p<0.05; ***p<0.01 |
|---|---|

**Table S11:** OLS summary of Win Percentage and eigenvectors in France. *Note: *p<0.1; **p<0.05; ***p<0.01.*

|  | Dependent variable: | | | | | |
|---|---|---|---|---|---|---|
|  | Divisiveness | | | | | |
|  | (1) | (2) | (3) | (4) | (5) | (6) |
| First Eigenvector | 0.334*** <br> p = 0.007 |  | 0.737*** <br> p = 0.00001 |  | 0.263** <br> p = 0.031 | 0.647*** <br> p = 0.0002 |
| Second Eigenvector |  | −0.064 <br> p = 0.615 | −0.578*** <br> p = 0.0003 |  |  | −0.514*** <br> p = 0.002 |
| Third Eigenvector |  |  |  | 0.346*** <br> p = 0.005 | 0.279** <br> p = 0.022 | 0.177 <br> p = 0.127 |
| Constant | −0.000 <br> p = 1.000 | −0.000 <br> p = 1.000 | −0.000 <br> p = 1.000 | −0.000 <br> p = 1.000 | −0.000 <br> p = 1.000 | −0.000 <br> p = 1.000 |
| Observations | 65 | 65 | 65 | 65 | 65 | 65 |
| $R^2$ | 0.112 | 0.004 | 0.283 | 0.120 | 0.184 | 0.310 |
| Adjusted $R^2$ | 0.098 | −0.012 | 0.260 | 0.106 | 0.158 | 0.276 |
| Residual Std. Error | 0.950 (df = 63) | 1.006 (df = 63) | 0.860 (df = 62) | 0.946 (df = 63) | 0.918 (df = 62) | 0.851 (df = 61) |
| F Statistic | 7.917*** (df = 1; 63) | 0.256 (df = 1; 63) | 12.243*** (df = 2; 62) | 8.582*** (df = 1; 63) | 7.011*** (df = 2; 62) | 9.149*** (df = 3; 61) |

**Table S12:** OLS summary of Divisiveness and eigenvectors in France. *Note:* *$p<0.1$; **$p<0.05$; ***$p<0.01$.

Note: *$p<0.1$; **$p<0.05$; ***$p<0.01$.

|  | Dependent variable: | | | | | |
|---|---|---|---|---|---|---|
|  | Win Percentage | | | | | |
|  | (1) | (2) | (3) | (4) | (5) | (6) |
| First Eigenvector | −0.994*** <br> p = 0.000 |  | −0.927*** <br> p = 0.000 |  | −0.996*** <br> p = 0.000 | −0.928*** <br> p = 0.000 |
| Second Eigenvector |  | 0.765*** <br> p = 0.000 | 0.092*** <br> p = 0.00000 |  |  | 0.092*** <br> p = 0.00000 |
| Third Eigenvector |  |  |  | 0.110 <br> p = 0.374 | −0.010 <br> p = 0.470 | −0.001 <br> p = 0.923 |
| Constant | 0.000 <br> p = 1.000 | 0.000 <br> p = 1.000 | 0.000 <br> p = 1.000 | 0.000 <br> p = 1.000 | 0.000 <br> p = 1.000 | 0.000 <br> p = 1.000 |
| Observations | 67 | 67 | 67 | 67 | 67 | 67 |
| $R^2$ | 0.989 | 0.585 | 0.993 | 0.012 | 0.989 | 0.993 |
| Adjusted $R^2$ | 0.989 | 0.579 | 0.993 | −0.003 | 0.989 | 0.993 |
| Residual Std. Error | 0.106 (df = 65) | 0.649 (df = 65) | 0.085 (df = 64) | 1.002 (df = 65) | 0.106 (df = 64) | 0.085 (df = 63) |
| F Statistic | 5,834.477*** (df = 1; 65) | 91.749*** (df = 1; 65) | 4,561.778*** (df = 2; 64) | 0.802 (df = 1; 65) | 2,896.393*** (df = 2; 64) | 2,994.128*** (df = 3; 63) |

*Note:* *p<0.1; **p<0.05; ***p<0.01

**Table S13:** OLS summary of Win Percentage and eigenvectors in Brazil. *Note:* *p<0.1; **p<0.05; ***p<0.01.

|  | Dependent variable: |  |  |  |  |  |
|---|---|---|---|---|---|---|
|  | Divisiveness |  |  |  |  |  |
|  | (1) | (2) | (3) | (4) | (5) | (6) |
| First Eigenvector | 0.367*** <br> p = 0.003 |  | 0.745*** <br> p = 0.00002 |  | 0.366*** <br> p = 0.003 | 0.754*** <br> p = 0.00002 |
| Second Eigenvector |  | −0.020 <br> p = 0.875 | 0.520*** <br> p = 0.002 |  |  | 0.528*** <br> p = 0.002 |
| Third Eigenvector |  |  |  | −0.056 <br> p = 0.653 | −0.012 <br> p = 0.920 | 0.037 <br> p = 0.739 |
| Constant | 0.000 <br> p = 1.000 | 0.000 <br> p = 1.000 | 0.000 <br> p = 1.000 | 0.000 <br> p = 1.000 | 0.000 <br> p = 1.000 | 0.000 <br> p = 1.000 |
| Observations | 67 | 67 | 67 | 67 | 67 | 67 |
| R$^2$ | 0.135 | 0.0004 | 0.263 | 0.003 | 0.135 | 0.265 |
| Adjusted R$^2$ | 0.122 | −0.015 | 0.240 | −0.012 | 0.108 | 0.230 |
| Residual Std. Error | 0.937 (df = 65) | 1.007 (df = 65) | 0.872 (df = 64) | 1.006 (df = 65) | 0.944 (df = 64) | 0.878 (df = 63) |
| F Statistic | 10.133*** (df = 1; 65) | 0.025 (df = 1; 65) | 11.435*** (df = 2; 64) | 0.204 (df = 1; 65) | 4.995*** (df = 2; 64) | 7.555*** (df = 3; 63) |
| *Note:* |  |  |  |  |  | *p<0.1; **p<0.05; ***p<0.01 |

**Table S14:** OLS summary of Divisiveness and eigenvectors in Brazil. *Note: *p<0.1; **p<0.05; ***p<0.01.*

# 4 Supplementary Figures

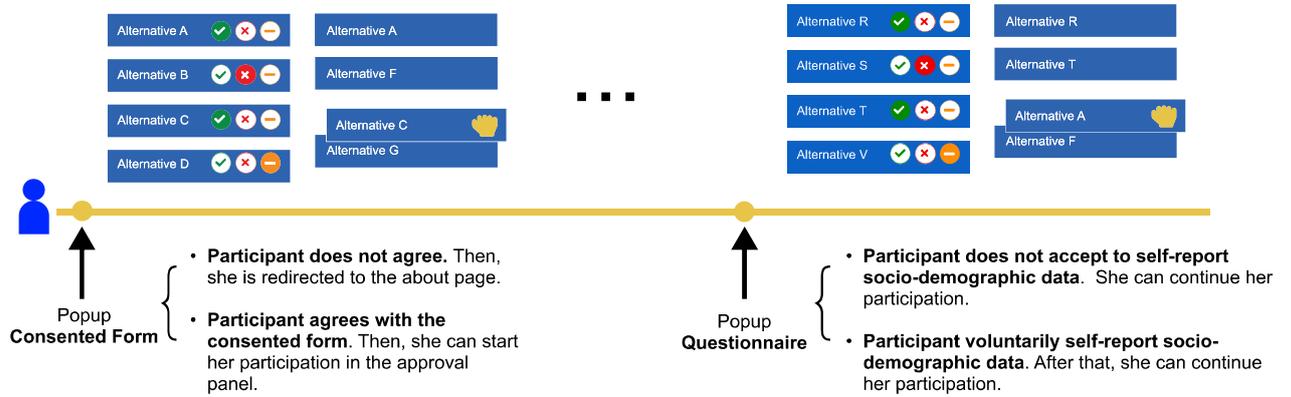

**Figure S1:** Life cycle of participation in the platform. It should be noted that the participant can finish her participation at any time.

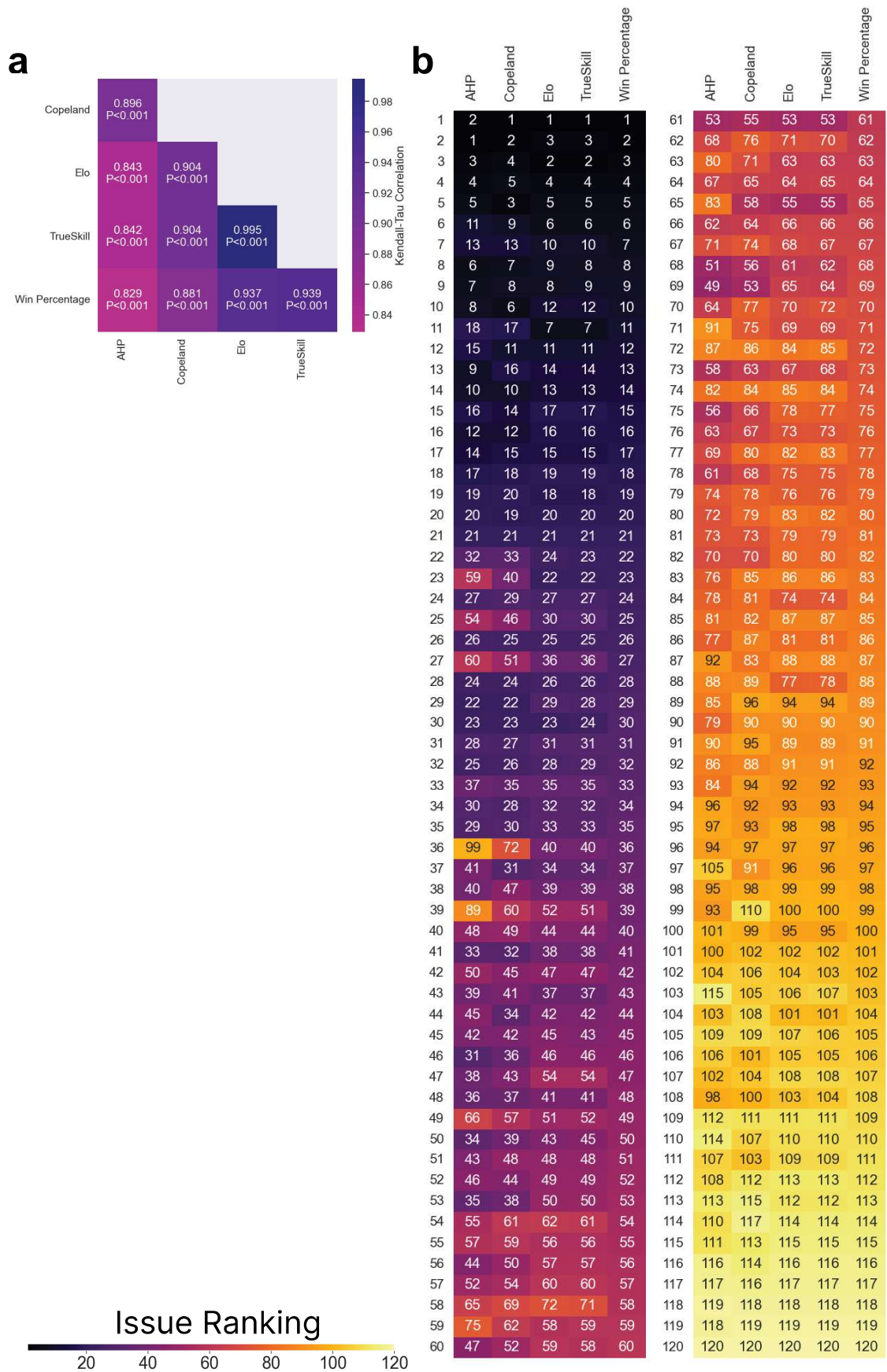

**Figure S2:** Robustness between rankings of agreements by aggregation functions (b) and the Kendall-Tau correlation matrix (a) in France. Win Percentage represents the ranking of agreements presented in the manuscript.

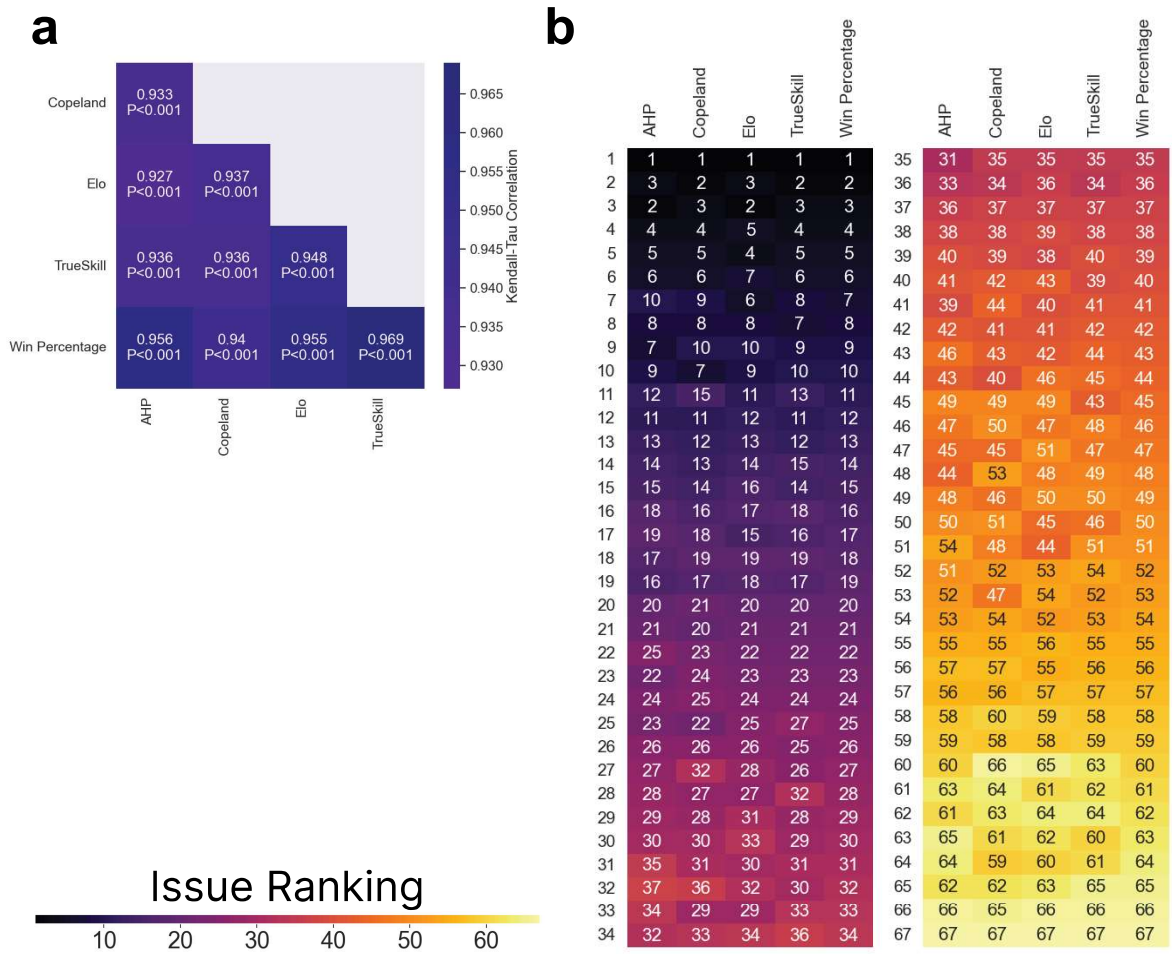

**Figure S3:** Robustness between rankings of agreements by aggregation functions (b) and the Kendall-Tau correlation matrix (a) in Brazil. Win Percentage represents the ranking of agreements presented in the manuscript.

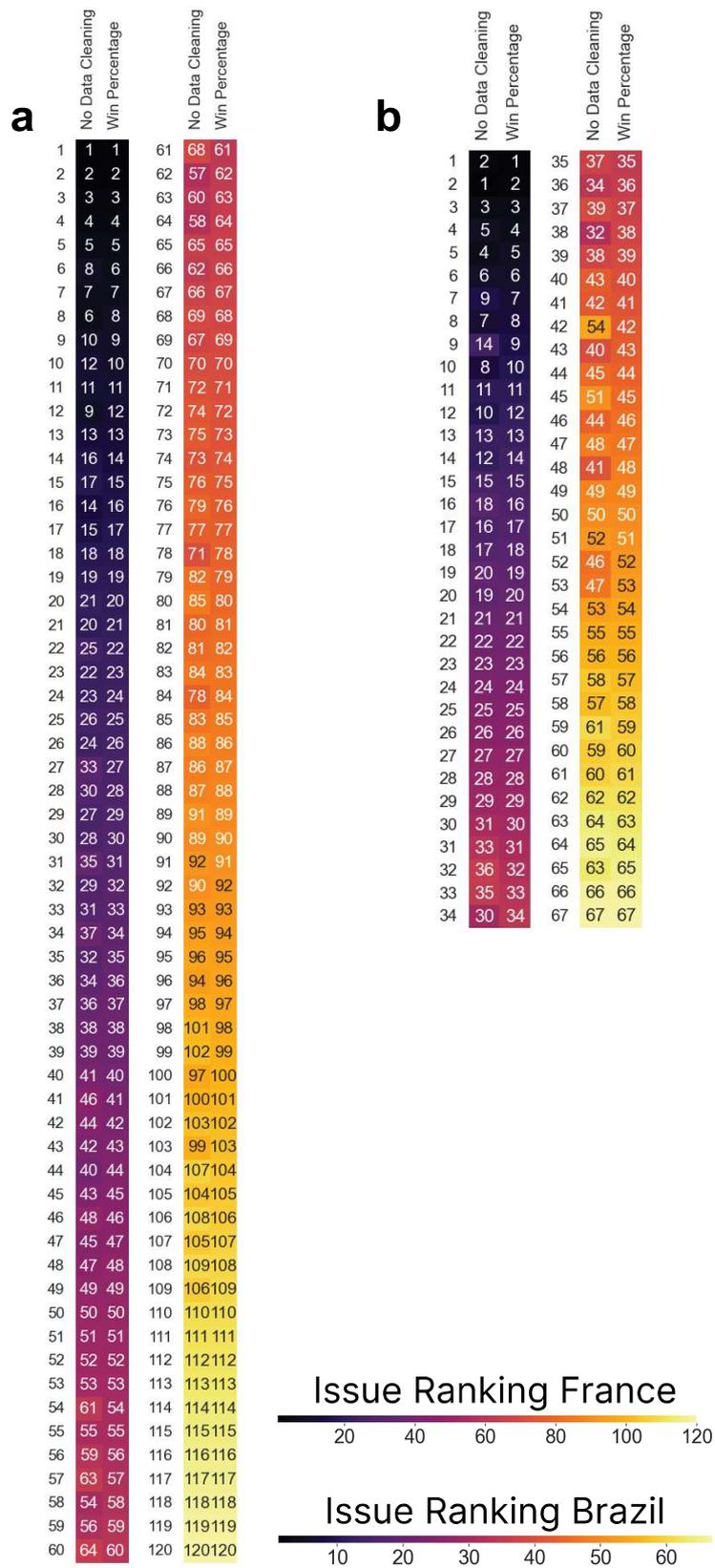

**Figure S4:** Robustness between rankings of agreements including the suspicious/bot participation in (a) France and (b) Brazil. Win Percentage represents the ranking of agreements presented in the manuscript.

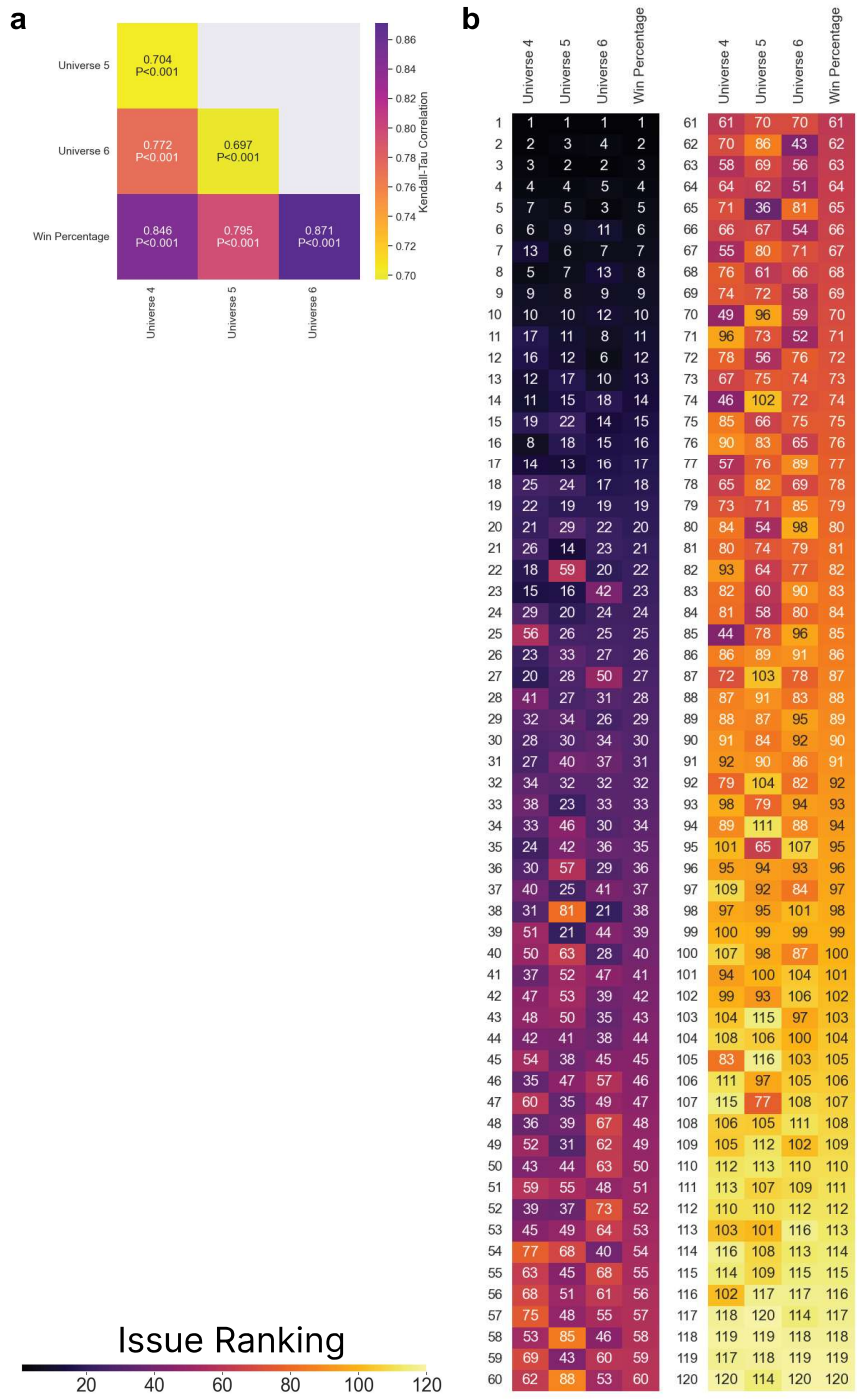

**Figure S5:** Robustness between the rankings of agreements across universes (b) and the Kendall-Tau correlation matrix (a) in France. Win Percentage represents the ranking of agreements presented in the manuscript.

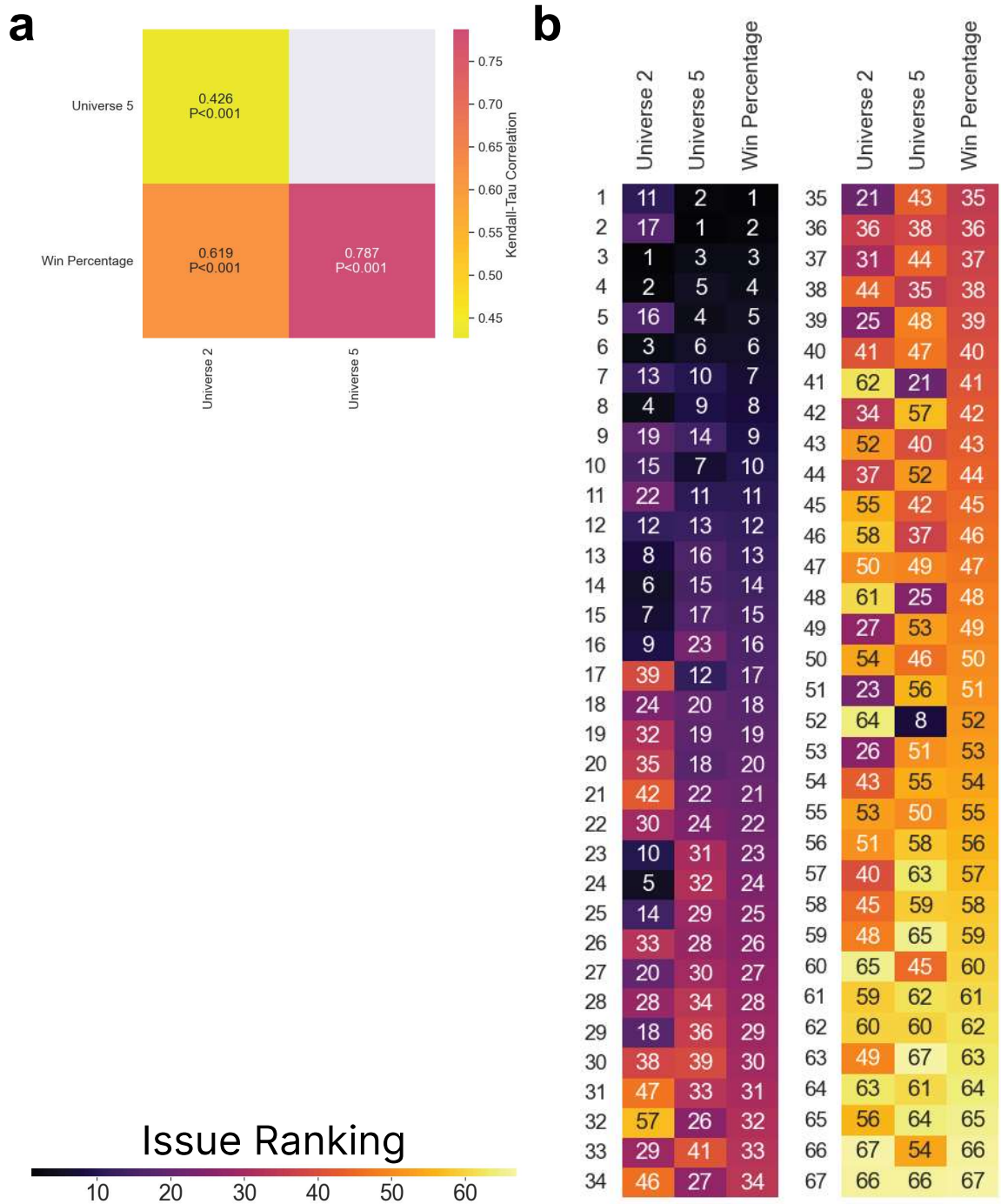

**Figure S6:** Robustness between the rankings of agreements across universes (b) and the Kendall-Tau correlation matrix (a) in Brazil. Win Percentage represents the ranking of agreements presented in the manuscript.

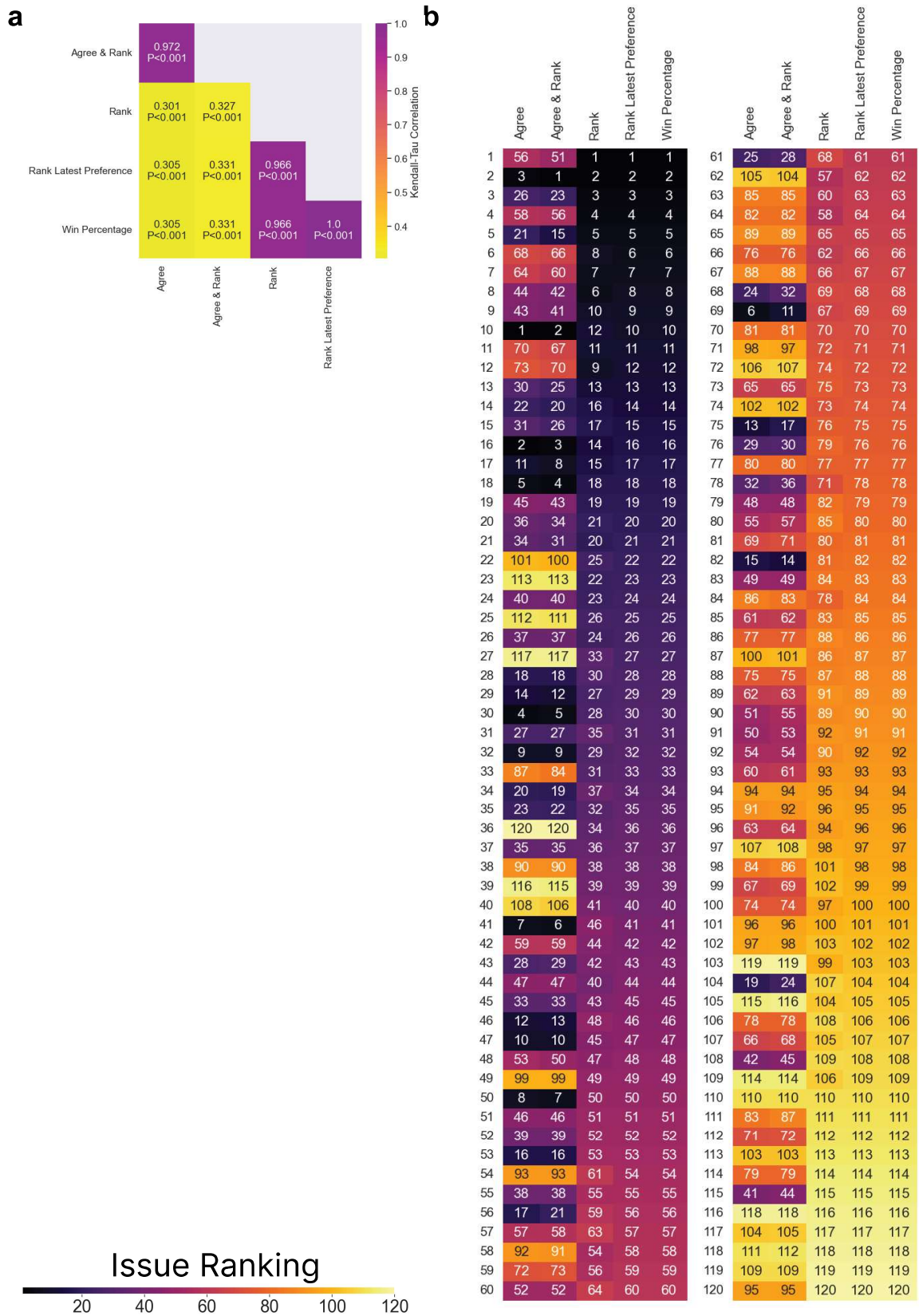

**Figure S7:** Robustness between the rankings of agreements using different datasets (a) and the Kendall-Tau correlation matrix (b). The manuscript uses the Win Percentage ranking of agreements.

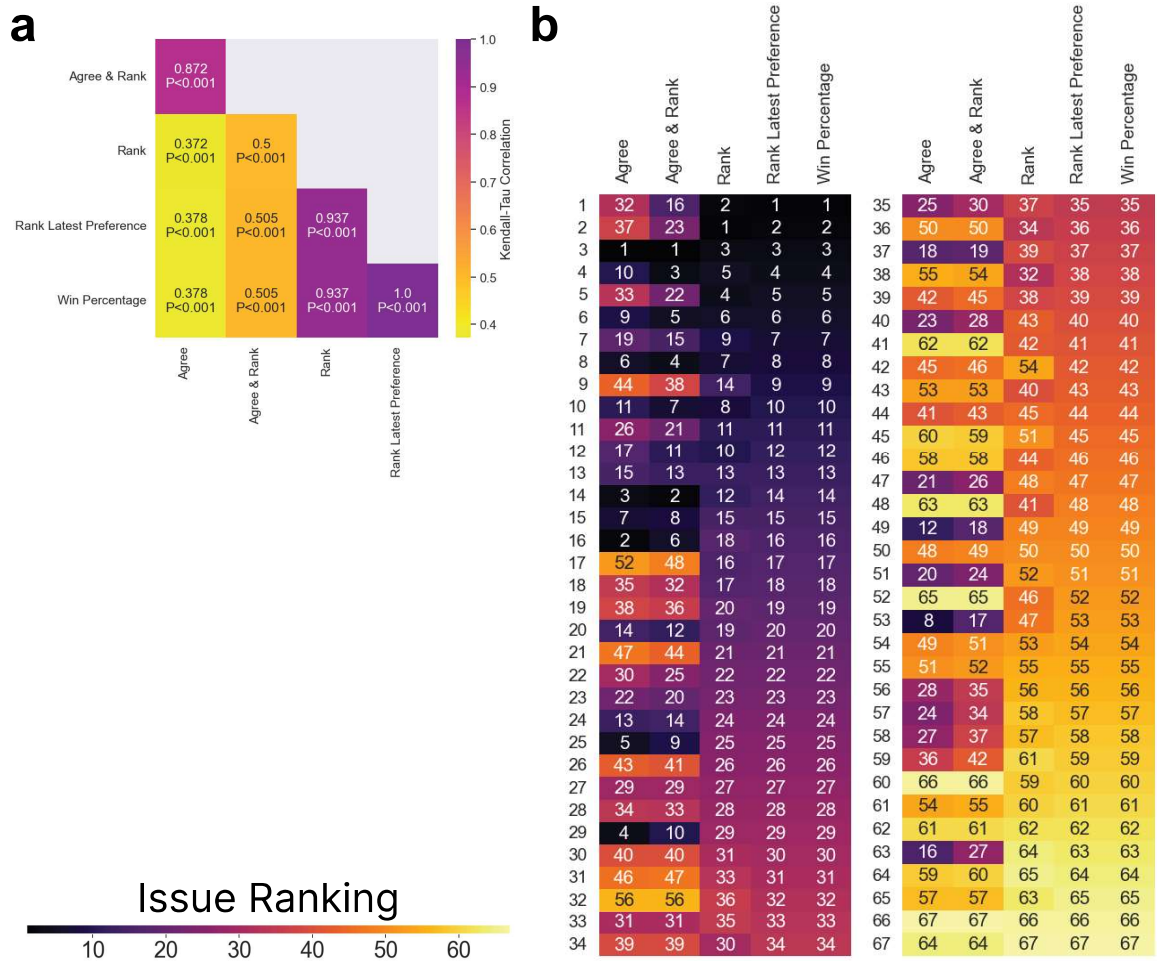

**Figure S8:** Robustness between the rankings of agreements using different datasets (a) and the Kendall-Tau correlation matrix (b) in Brazil. The manuscript uses the Win Percentage ranking of agreements.

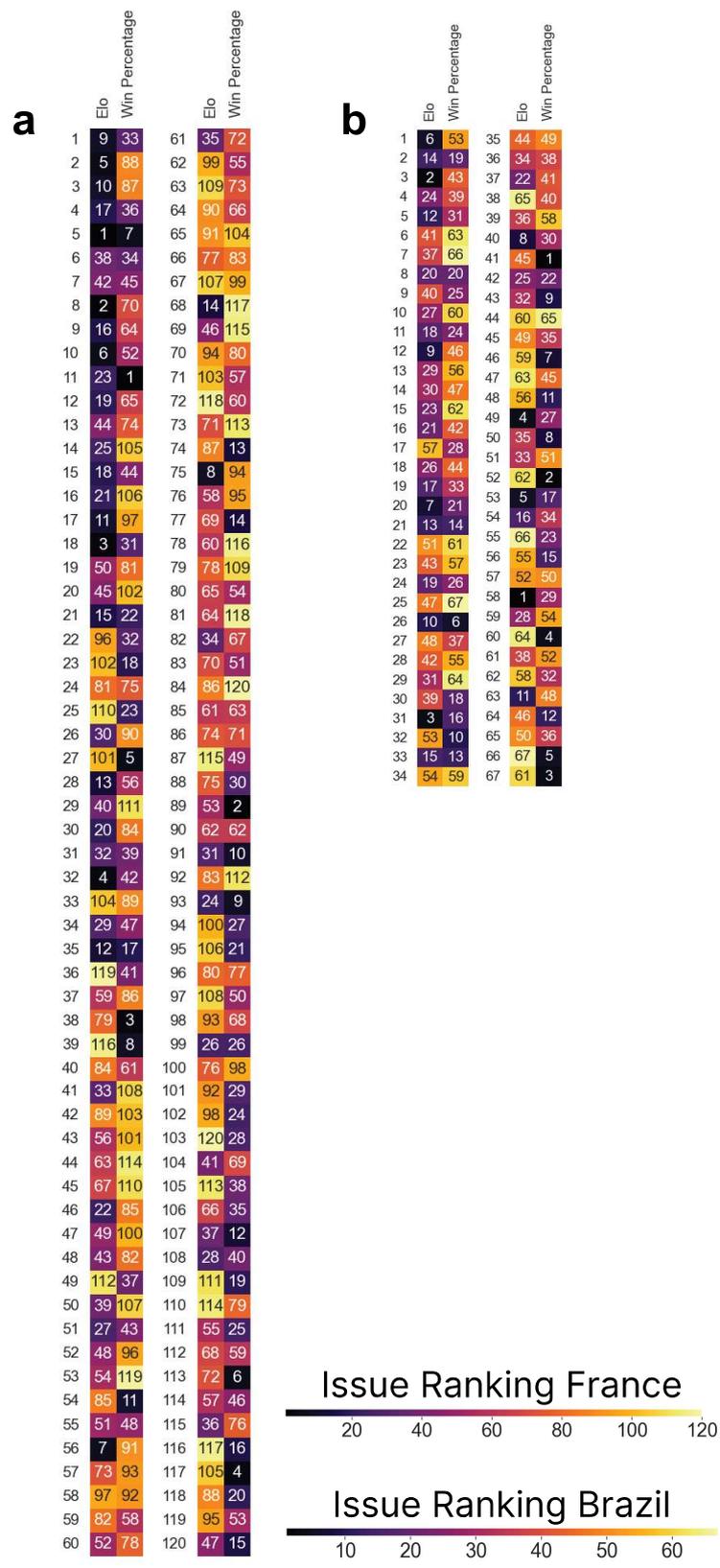

**Figure S9:** Robustness ranking of divisiveness comparing the results using Elo and Win Percentage in (a) France and (b) Brazil. The manuscript presents the results using the divisiveness ranking of Win Percentage.

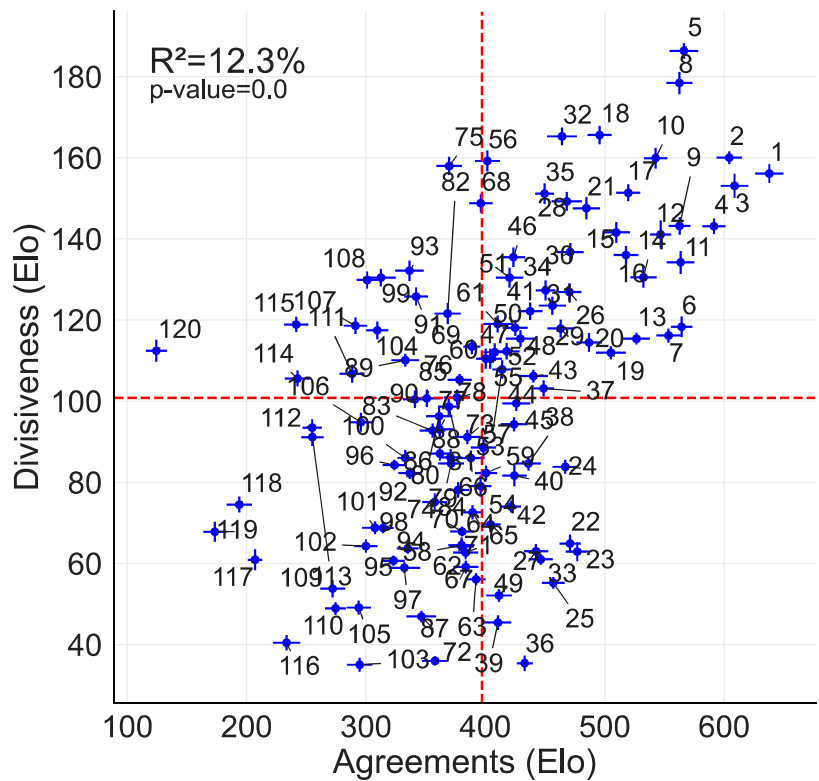

**Figure S10:** Agreements and divisiveness using Elo score method in France. Each point represents the mean score of a proposal, and the error bars represent the 95% confidence interval of the proposal's score. Both values were calculated by bootstrapping half of the dataset 30 times. We report the $R^2$ calculated as the square of Pearson's correlation estimated from a two-sided alternative hypothesis, as determined by the SciPy library (v.1.9.3).

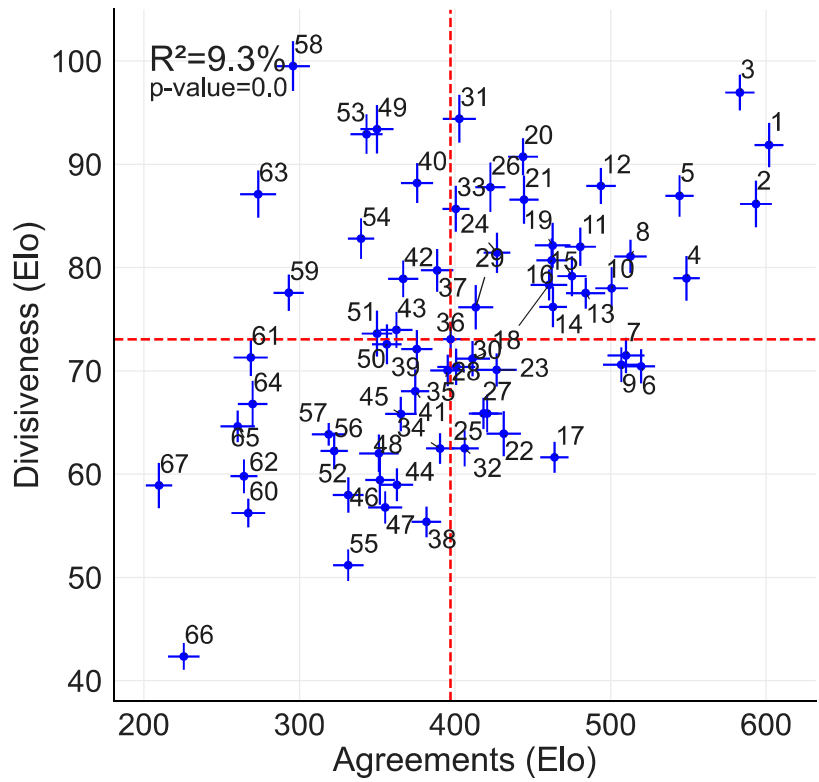

**Figure S11:** Agreements and divisiveness using Elo score method in Brazil. Each point represents the mean score of a proposal, and the error bars represent the 95% confidence interval of the proposal's score. Both values were calculated by bootstrapping half of the dataset 30 times. We report the $R^2$ calculated as the square of Pearson's correlation estimated from a two-sided alternative hypothesis, as determined by the SciPy library (v.1.9.3).

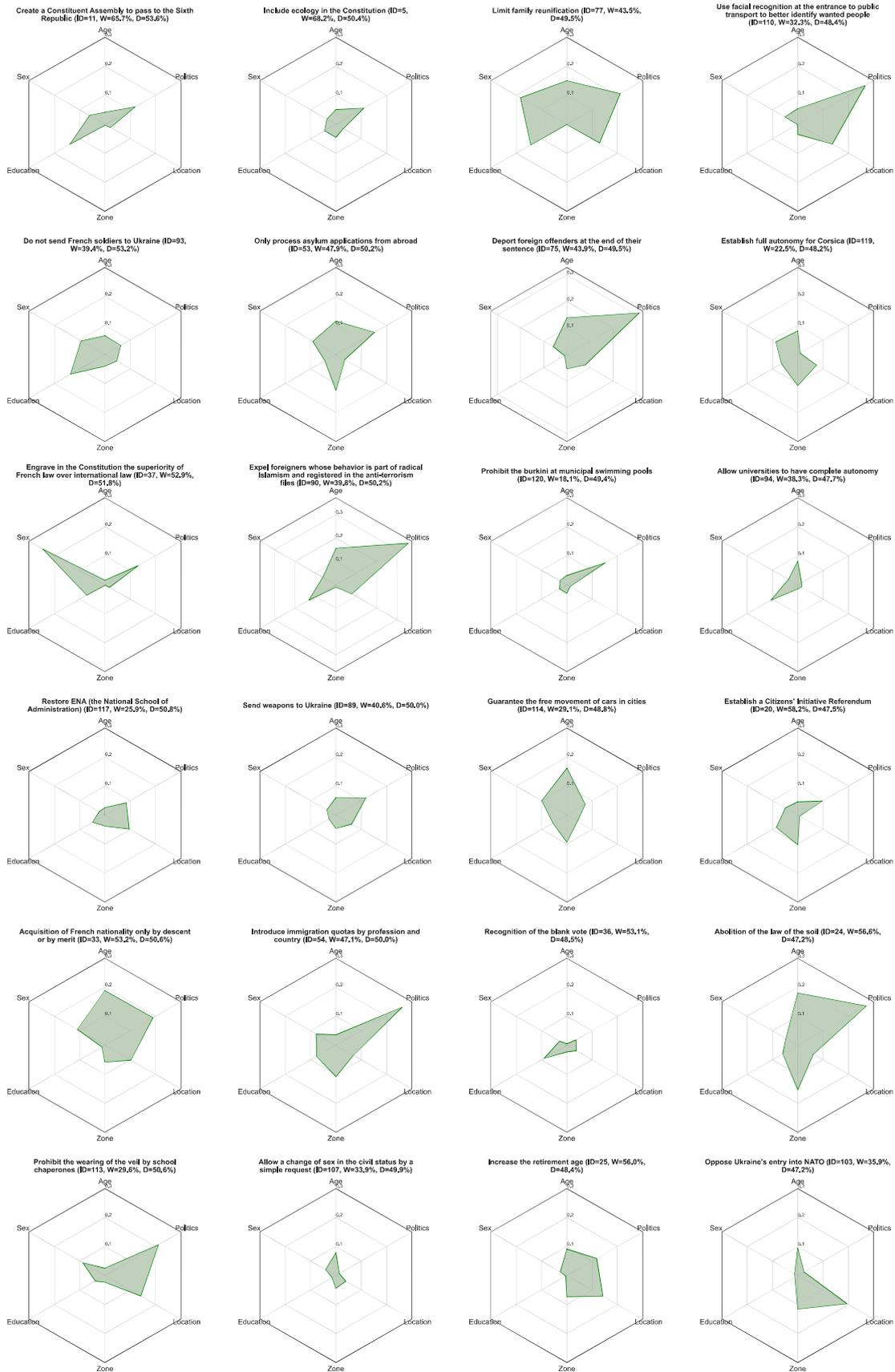

**Figure S12:** Multidimensional divisiveness of issues in France. Proposals were sorted from the most to the least divisive. Part 1 of 5.

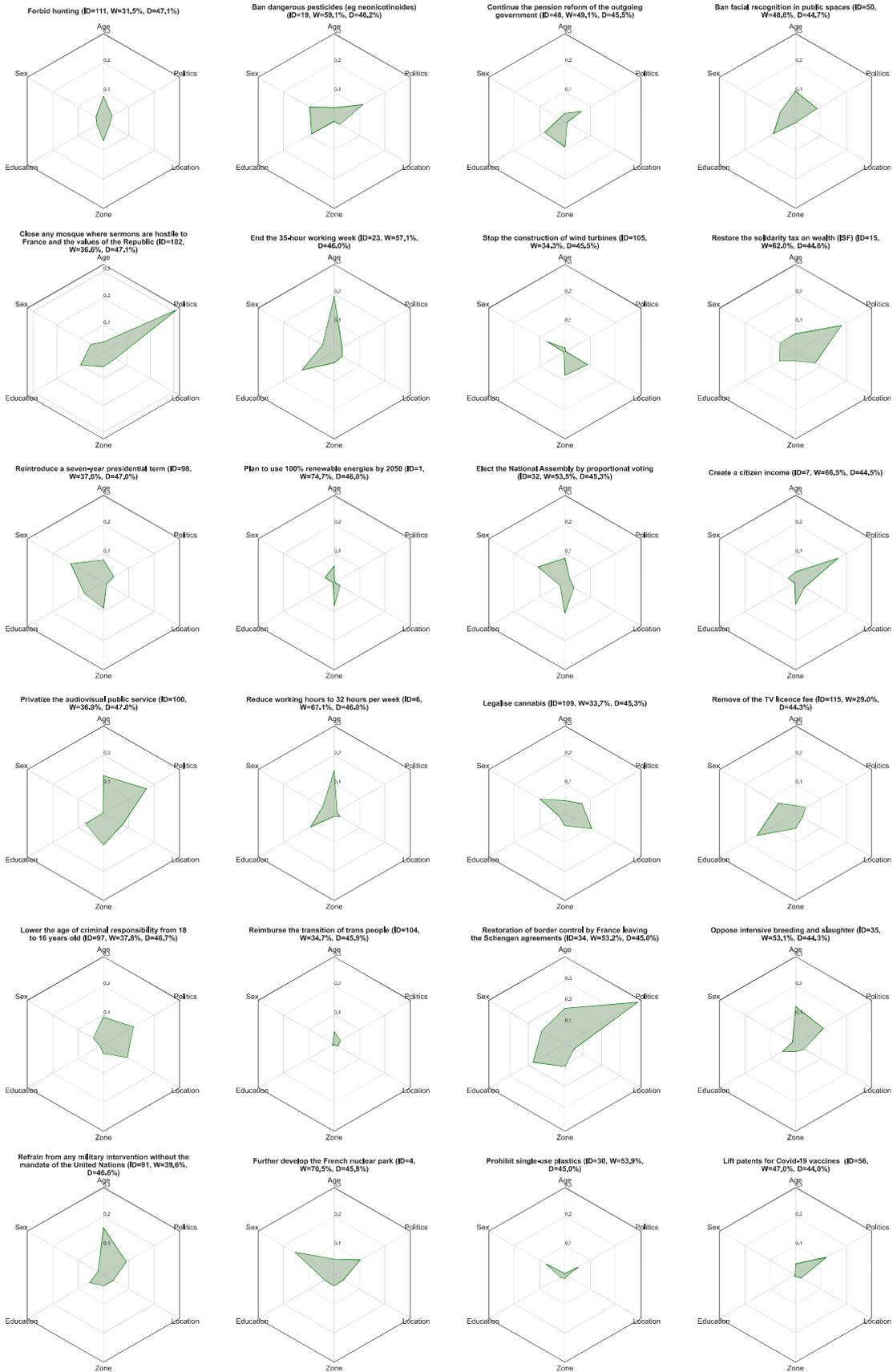

**Figure S13:** Multidimensional divisiveness of issues in France. Proposals were sorted from the most to the least divisive. Part 2 of 5.

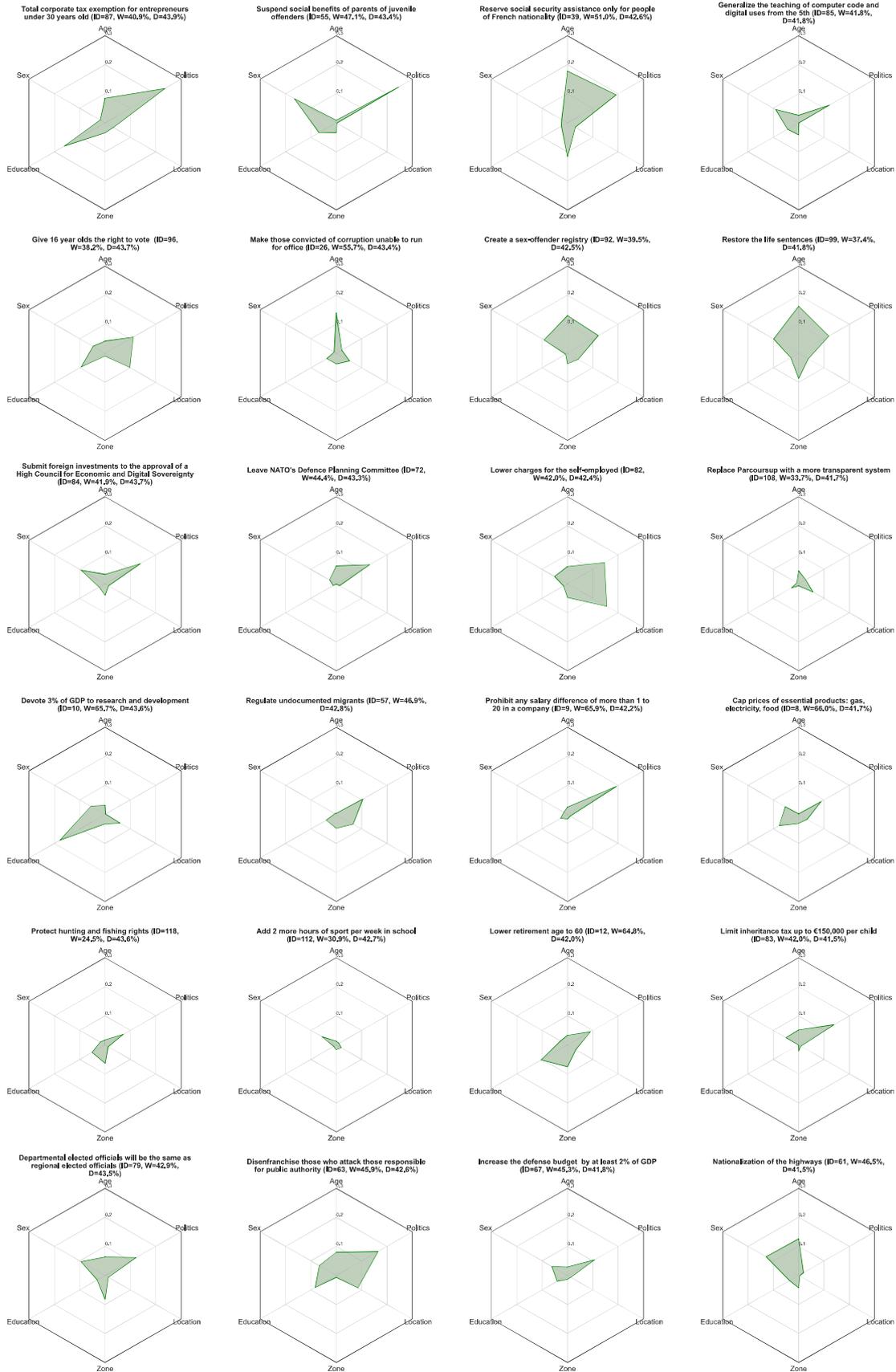

**Figure S14:** Multidimensional divisiveness of issues in France. Proposals were sorted from the most to the least divisive. Part 3 of 5.

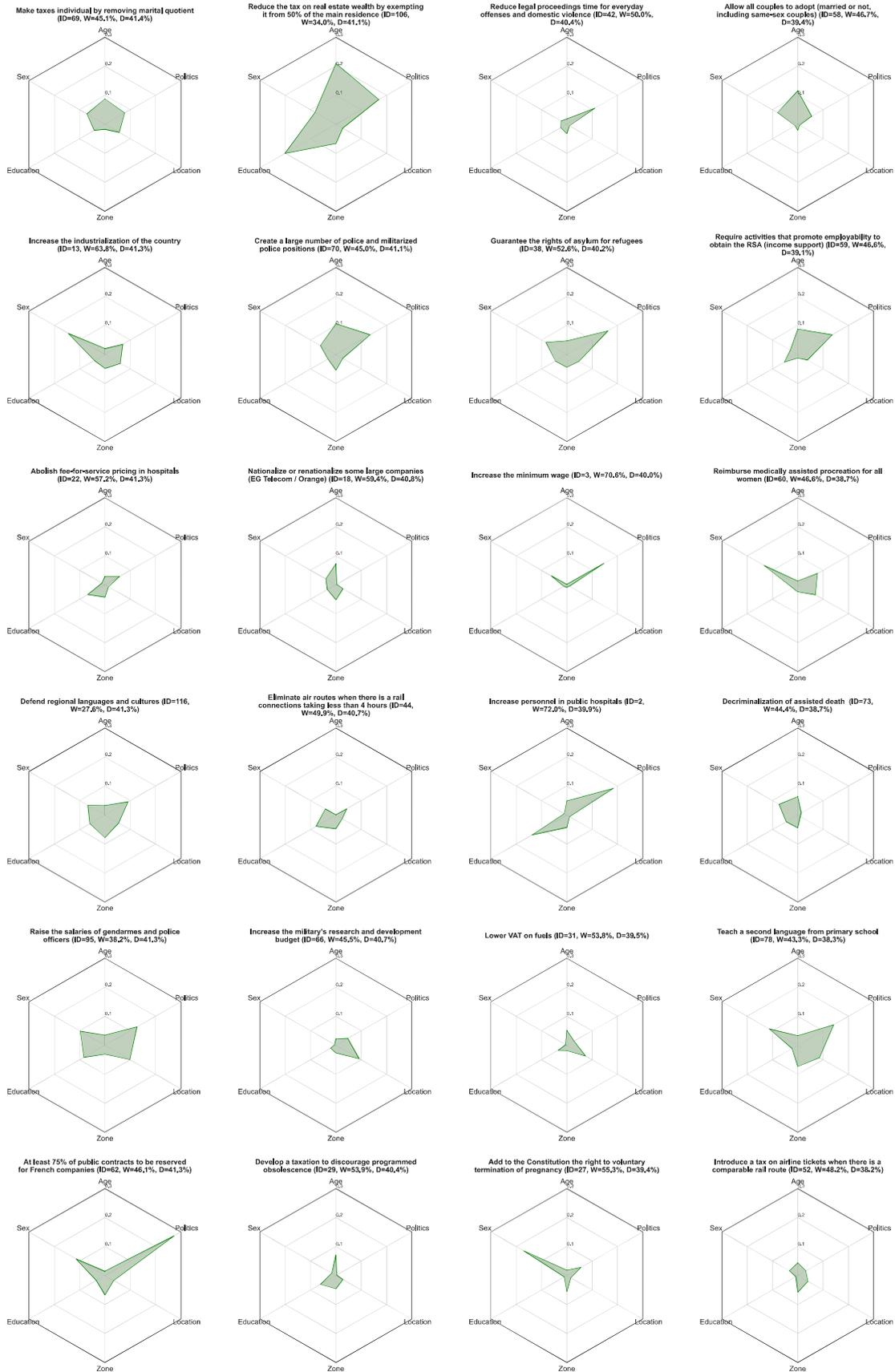

**Figure S15:** Multidimensional divisiveness of issues in France. Proposals were sorted from the most to the least divisive. Part 4 of 5.

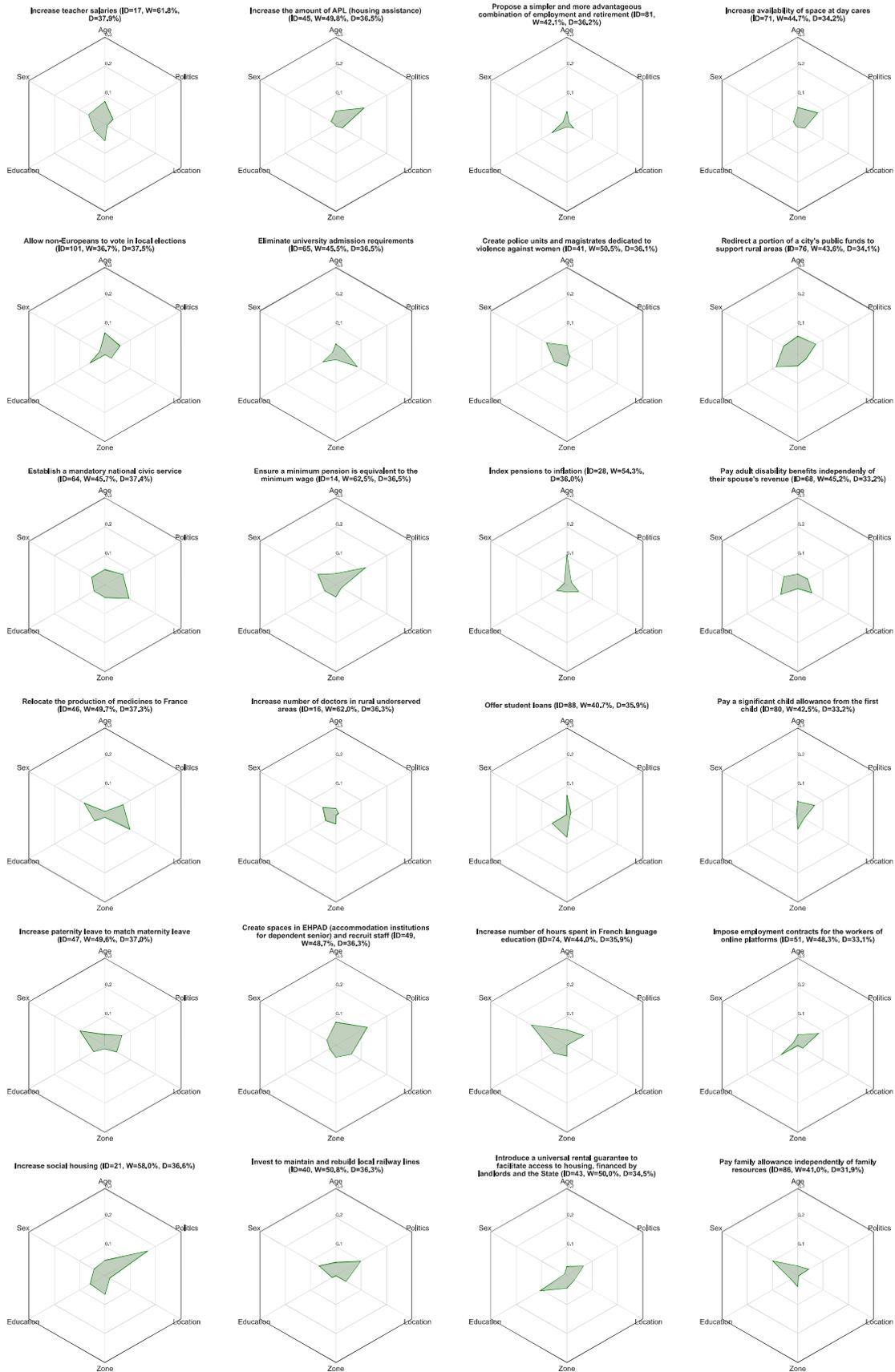

**Figure S16:** Multidimensional divisiveness of issues in France. Proposals were sorted from the most to the least divisive. Part 5 of 5.

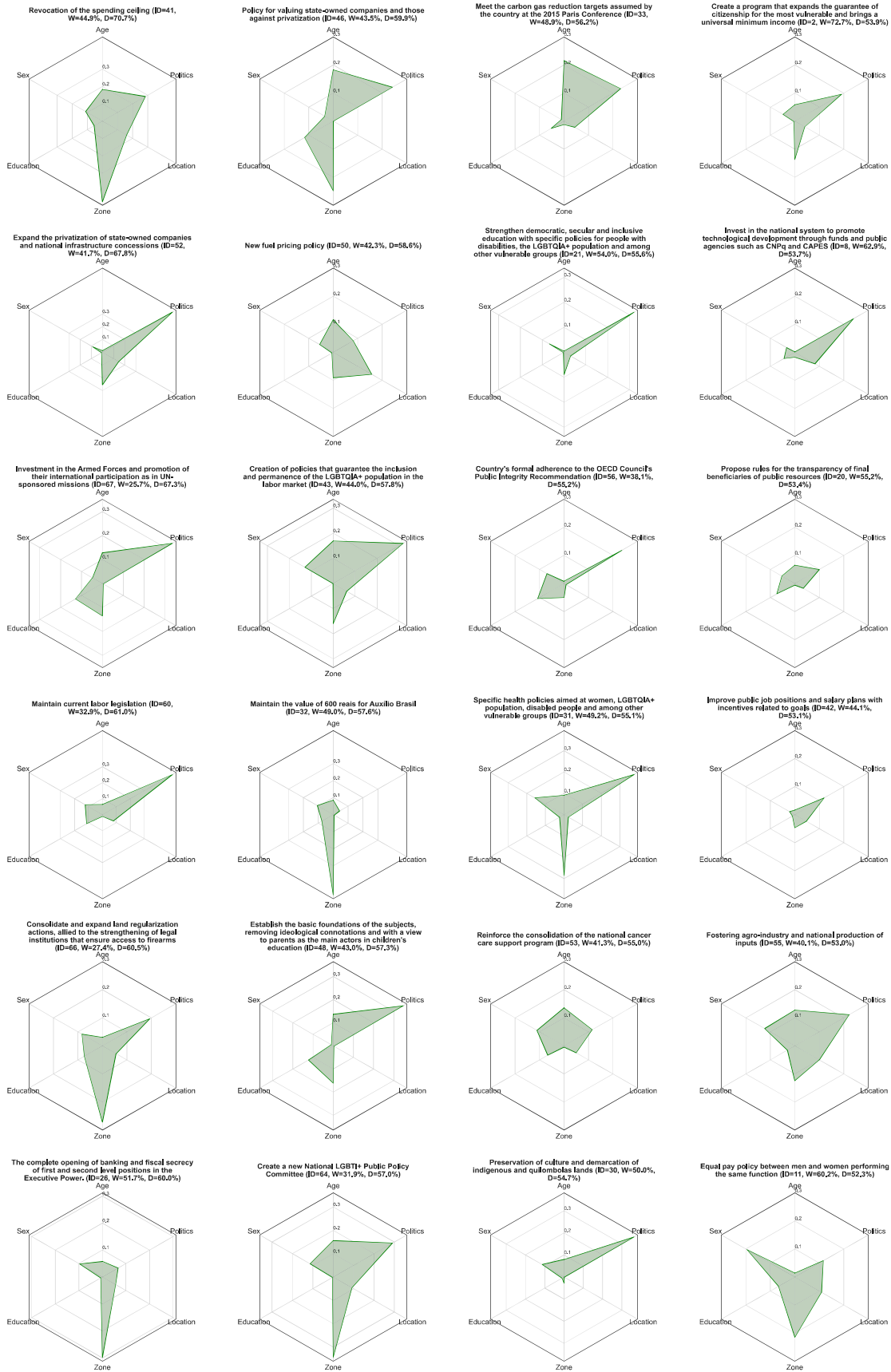

**Figure S17:** Multidimensional divisiveness of issues in Brazil. Proposals were sorted from the most to the least divisive. Part 1 of 3.

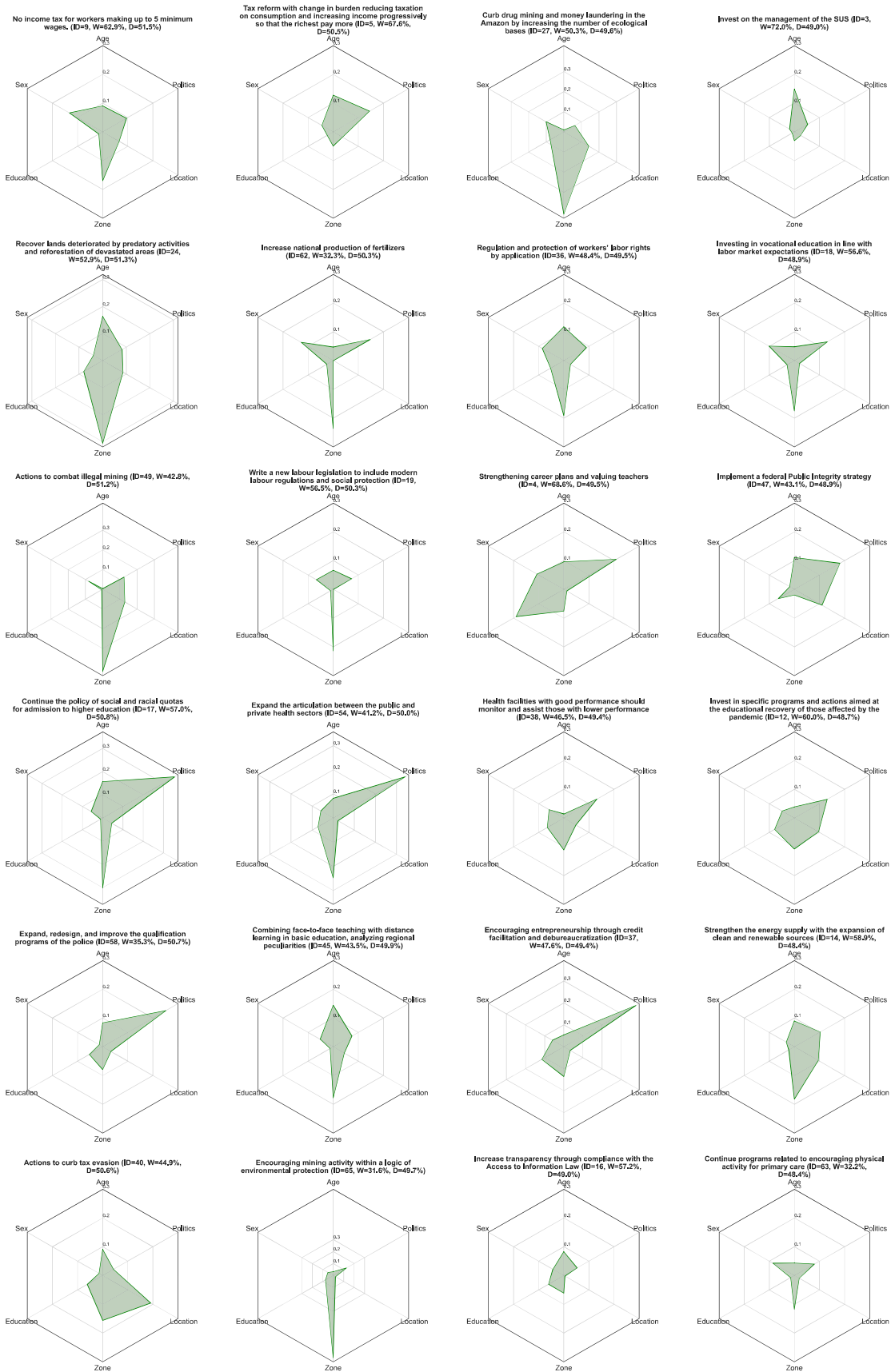

**Figure S18:** Multidimensional divisiveness of issues in Brazil. Proposals were sorted from the most to the least divisive. Part 2 of 3.

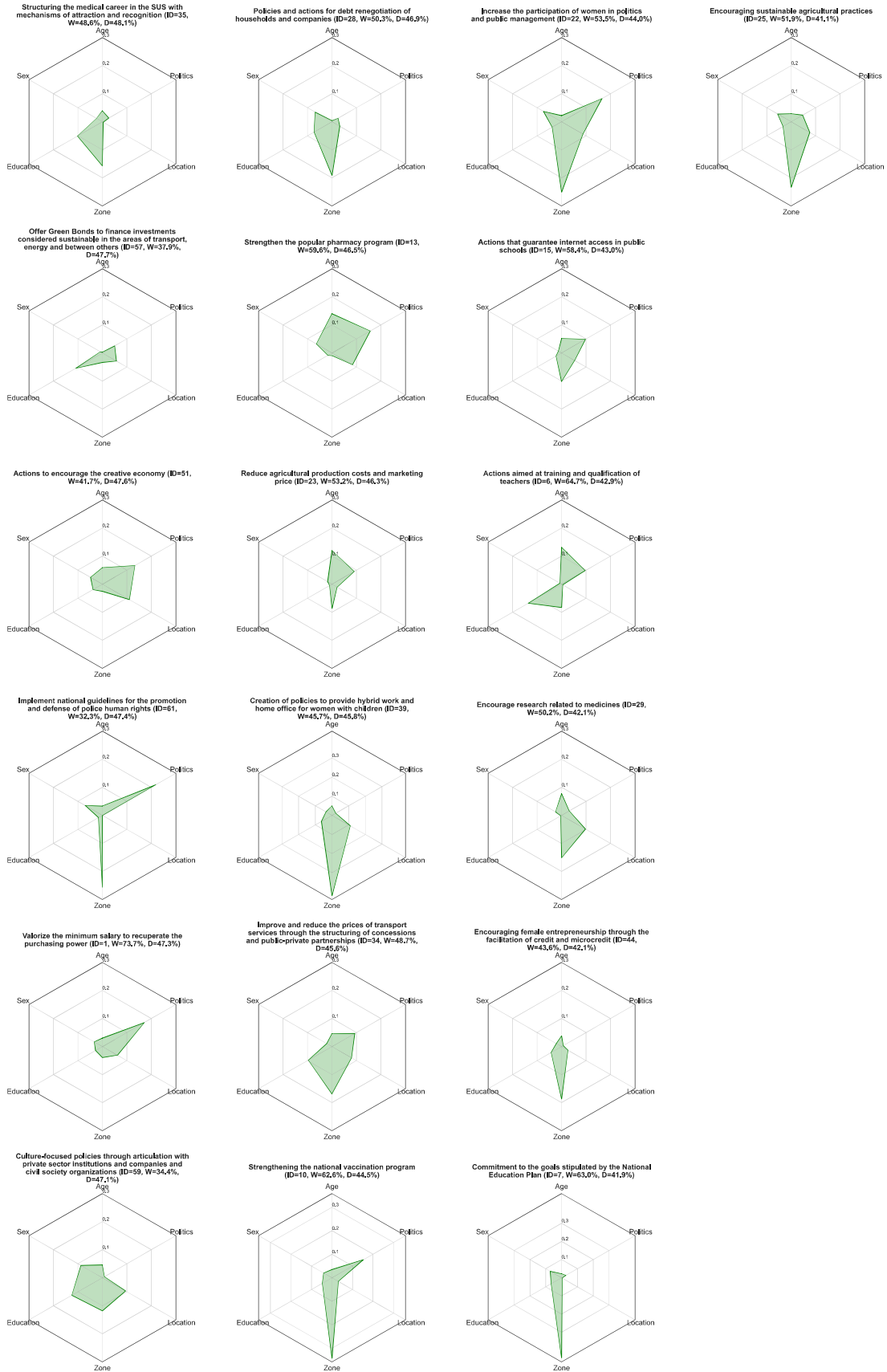

**Figure S19:** Multidimensional divisiveness of issues in Brazil. Proposals were sorted from the most to the least divisive. Part 3 of 3.

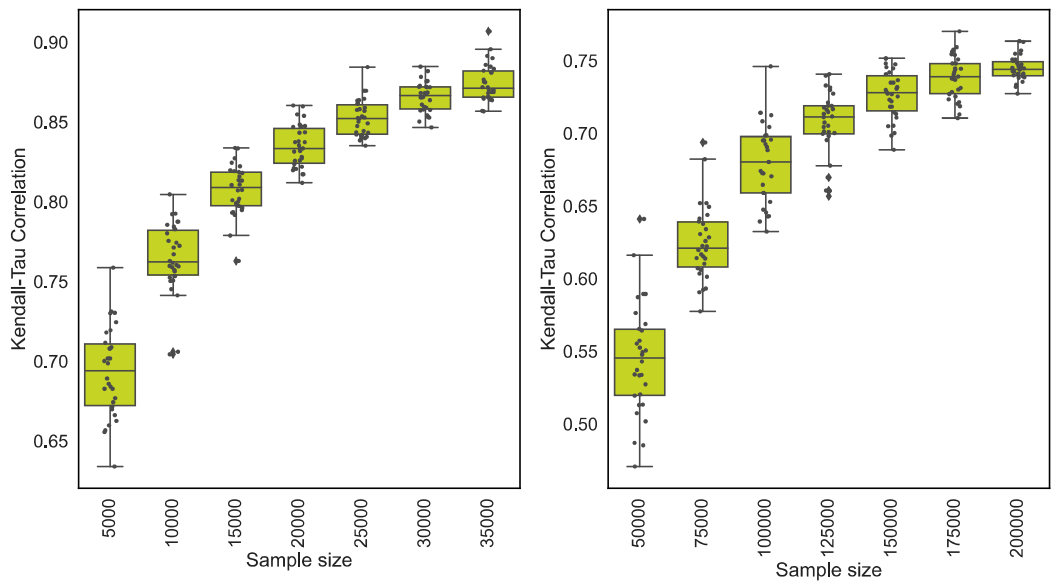

**Figure S20:** Convergence of values for (a) Agreements and (b) divisiveness by sampling the data set (y-axis). The scale in the x-axis starts in 5000, with steps of 5000. In the case of divisiveness, the x-axis starts in 50.000, with steps of 25.000. Boxplot figures: center lines show the medians; box limits indicate the 25th and 75th percentiles as determined by the seaborn library (v.0.12.1); whiskers extend 1.5 times the interquartile range from the 25th and 75th percentiles, and circles represent individual data points.

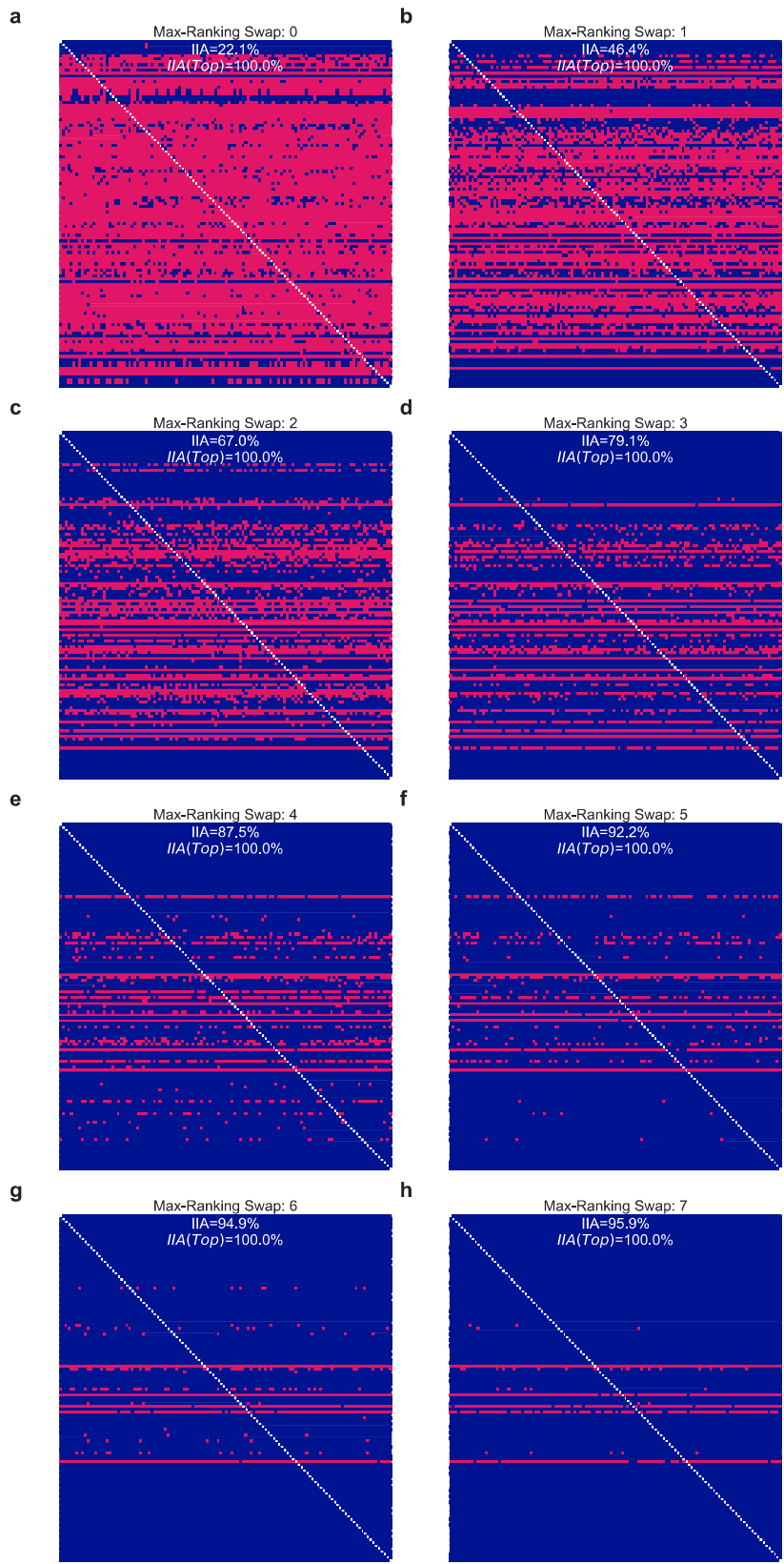

**Figure S21:** Independence of Irrelevant Alternatives (IIA) by allowing $k$ swaps in Ranking of Agreements in France.

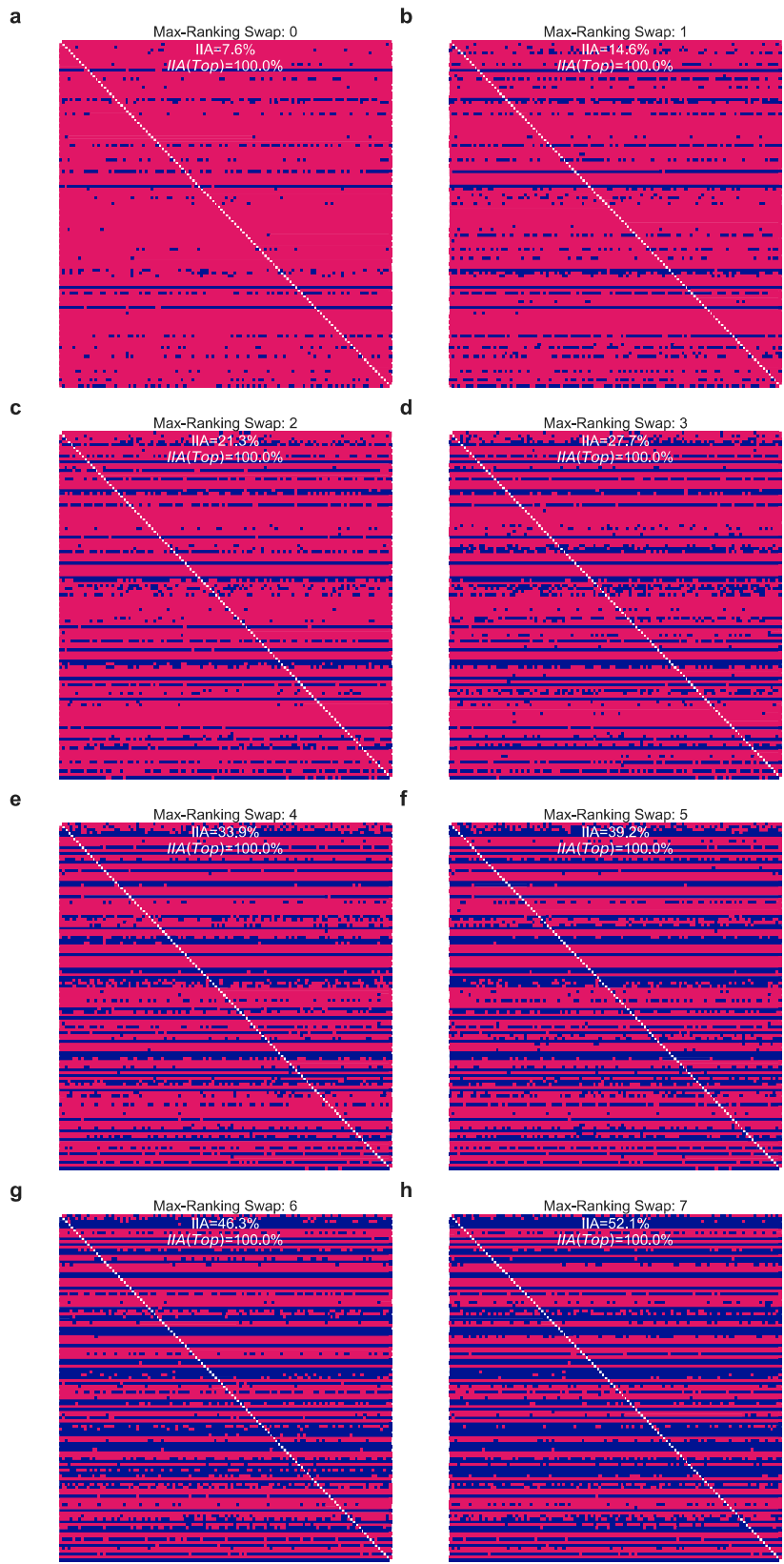

**Figure S22:** Independence of Irrelevant Alternatives (IIA) by allowing $k$ swaps in Ranking of divisiveness in France.

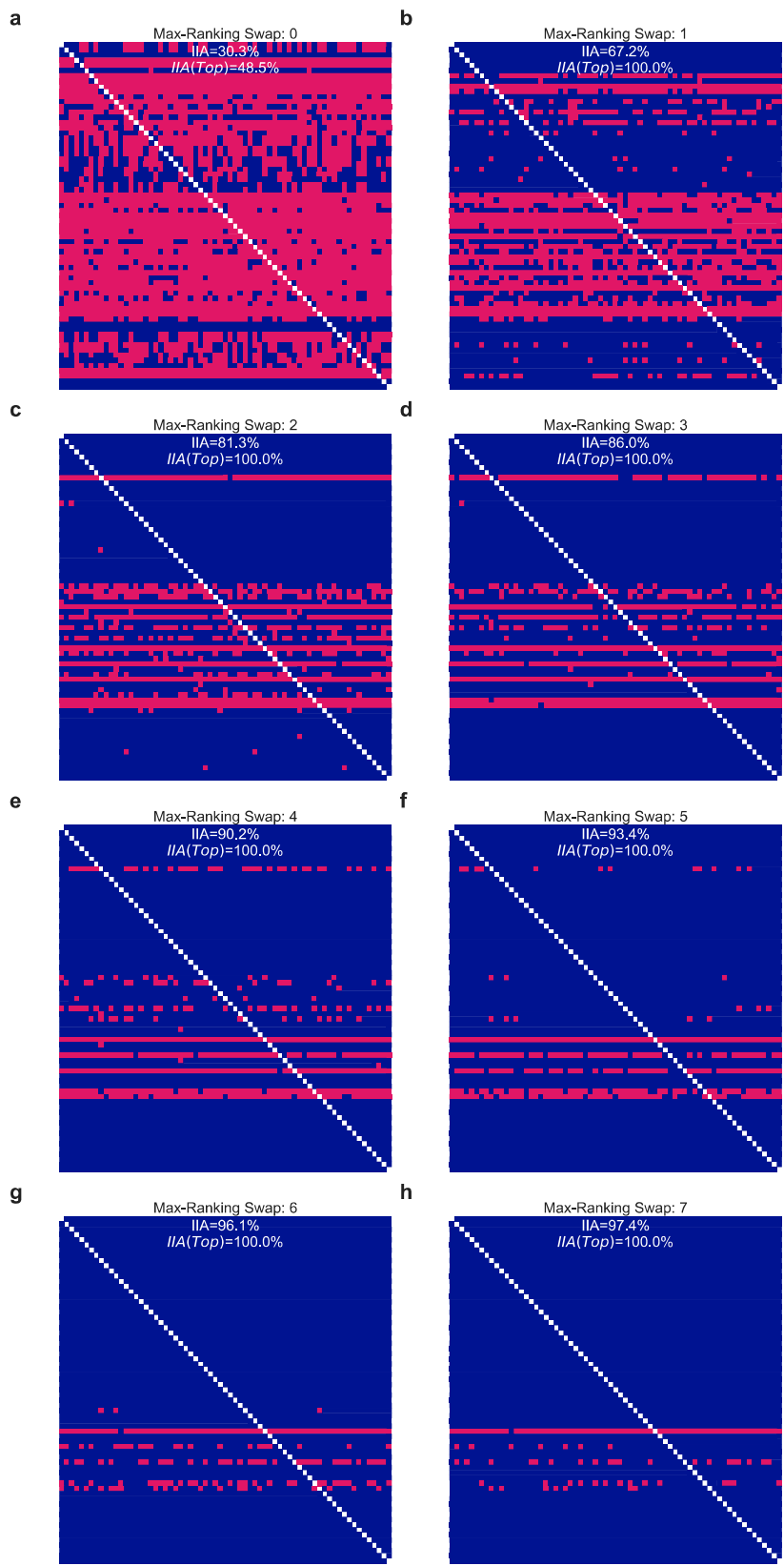

**Figure S23:** Independence of Irrelevant Alternatives (IIA) by allowing $k$ swaps in ranking of Agreements in Brazil.

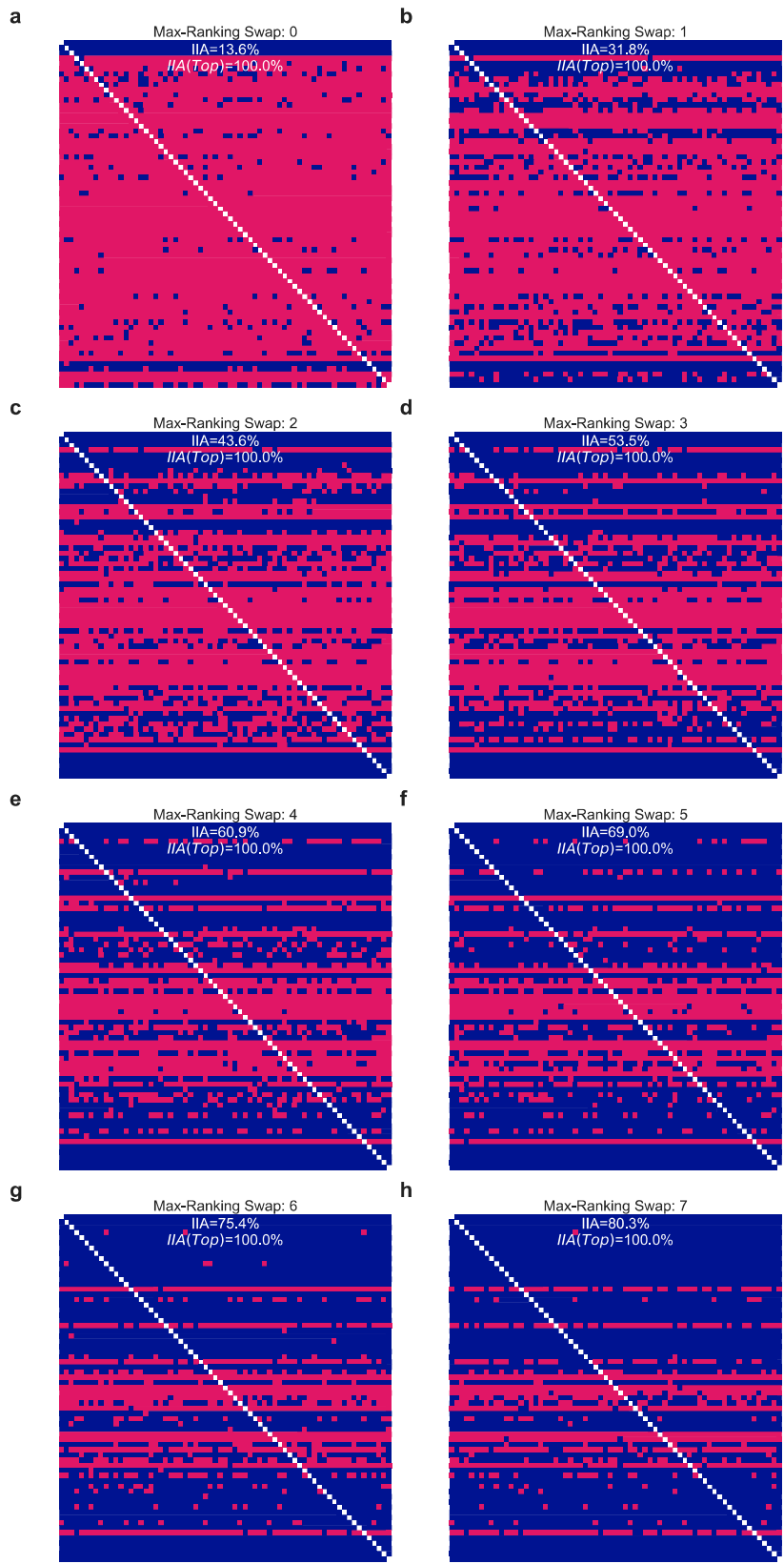

**Figure S24:** Independence of Irrelevant Alternatives (IIA) by allowing $k$ swaps in ranking of divisiveness in Brazil.